\documentclass{article}

\usepackage{graphicx} % Required for inserting images
\usepackage{amssymb}
\usepackage{algorithmic}
\usepackage{algorithm}
\usepackage{amsmath}
\usepackage{xcolor}
\usepackage{enumitem}
\usepackage{booktabs}
\usepackage{multirow}
\usepackage{makecell}
\newcommand{\cA}{\cal A}
\newcommand{\OPT}{\mathrm{OPT}}
\newcommand{\FOPT}{\mathrm{FOPT}}

\newcommand{\pij}{p_{ij}}
\newcommand{\xij}{x_{ij}}

\newcommand{\bx}{\bar{\mathbf{x}}}
\newcommand{\bxh}{\bar{\mathbf{x}}_H}
\newcommand{\byo}{\bar{\mathbf{y}}_1}
\newcommand{\byt}{\bar{\mathbf{y}}_2}
\newcommand{\bxhath}{\bar{\mathbf{x}}_{\hat{H}}}

\newcommand{\cmaxs}{C_{max}^*}
\newcommand{\cmins}{C_{min}^*}
\newcommand{\intpos}{{\mathbb{Z}}^+}

\newcommand{\bml}{M\setminus M_L}

\newcommand{\wtlp}{\mathbf{LP}_{\mathsf{LM}}}

\newcommand{\hatlp}{\widehat{\mathbf{LP}}}
\newcommand{\hatlplm}{\widehat{\mathbf{LP}_{\mathsf{LM}}}}
\newcommand{\bh}{\bar{H}}

\newcommand{\hati}{\hat{i}}

\newcommand{\ml}{M_L}

\newcommand{\xev}{x_{ev}}
\newcommand{\eob}{E^L}
\newcommand{\MM}{\sc Min Makespan}
\newcommand{\GB}{\sc Graph Balancing}
\newcommand{\MMGB}{\sc Max-Min Graph Balancing}
\newcommand{\SC}{\sc Santa Claus}
\newcommand{\RA}{\sc Restricted Assignment}
\newcommand{\vo}{val_{\textnormal{O}}(I)}
\newcommand{\LPLS}{\textbf{LP}_{\mathsf{LS}}}

\newtheorem{theorem}{Theorem}[section]

\newtheorem{corollary}[theorem]{Corollary}
\newtheorem{lemma}[theorem]{Lemma}
\newtheorem{claim}[theorem]{Claim}

                      {}
\def\eod{\vrule height 6pt width 5pt depth 0pt}
\newenvironment{proof}{\noindent {\bf Proof:}}
{\hfill\eod}

\title{
Trusted Advice for Scheduling on Unrelated Machines}
\author{Baruch Schieber \\ NJIT \\ Newark, NJ USA \\{\small \tt sbar@njit.edu} \and Hadas Shachnai \\ Technion \\Haifa, Israel \\{\tt \small hadas@cs.technion.ac.il} \and Tami Tamir \\ Reichman University \\ Herzliya, Israel \\{\small \tt tami@runi.ac.il} }
\date{}

\usepackage{lineno}

\begin{document}
\begin{titlepage}
\pagenumbering{gobble}
\maketitle

\begin{abstract}
Scheduling on unrelated parallel machines under standard objectives, such as makespan minimization or max–min fairness, has posed fundamental theoretical challenges for decades. A prominent example is the longstanding factor~$2$ approximation barrier for the {\sc Min Makespan} problem established by Lenstra, Shmoys, and Tardos (1990). We investigate these fundamental challenges within the framework of algorithms with trusted advice. We consider a setting in which a trusted oracle is accessed once and reveals partial but entirely accurate structural information about an optimal solution. This framework allows us to quantify the improvements in approximation guarantees enabled by various trusted advice.

We formalize a hierarchy of trusted advice oracles, ranging from a strong oracle that reveals the exact assignments of long jobs to a weaker oracle that indicates only which machines process such jobs. By exploiting this information, we obtain approximation guarantees beyond the classical worst-case barriers. In particular, we show that even the weakest oracle improves the approximation ratio for {\sc Min Makespan} to $\sqrt{3}\approx 1.732$. We also obtain improved ratios of $\frac{\sqrt{33}+1}{4}\approx 1.686$ and $\frac{5}{8}\sqrt{7}\approx 1.654$ for {\sc Restricted Assignment} and {\sc Graph Balancing}, respectively. For the max--min fairness objective, known as the {\sc Santa Claus} problem, we obtain a $2$-approximation using our weak oracle. Finally, we show that these problems remain NP-hard to approximate within certain constant factors even with the trusted advice oracles considered. Interestingly, for the {\sc Santa Claus} problem, we show a lower bound of 2 which is tight and matches the classical oracle-free bound.

\end{abstract}
\newpage
\setcounter{tocdepth}{2} % adjust to 1 if desired
%\tableofcontents

\end{titlepage}
\pagenumbering{arabic}
\setcounter{page}{1}

%\begin{document}
%\date{}
%\maketitle
%
%\input{abstract}

%\input{intro}
\section{Introduction}
\label{sec:intro}

Scheduling on unrelated parallel machines under standard objectives, such as makespan minimization or max–min fairness, poses major challenges in combinatorial optimization. For decades, classical approximation algorithms for these problems have encountered persistent theoretical barriers.
%Scheduling on unrelated parallel machines ($R\vert{}\vert{}C_{max}$) as well as several of its variants and generalizations (Generalized assignment, and the Max-Min Fairness (Santa Claus) problem) are 
A prominent example is the {\sc Min Makespan} problem, denoted in standard scheduling notation by $R||C_{\max}$. The input is a set of $n$ jobs, a set of $m$ machines, and an $m \times n$-matrix $p$, where $p_{ij}>0$ is the processing time of job $j$ on machine $i$.  The goal is to schedule all the jobs on the machines so as to minimize the latest completion time of any machine (i.e., the makespan).  
The classical linear programming (LP) rounding technique of Lenstra, Shmoys, and Tardos~\cite{LST90} achieves a $2$-approximation, a bound that remains notoriously difficult to break. Well explored classes of this problem are the {\sc Restricted Assignment} problem 
where $p_{ij}\in \{p_j, \infty \}$, and its special case the {\GB} problem, where each job can be assigned to at most two machines. Another well studied variant is the max-min fair allocation (known as the {\SC} problem), where the goal is to maximize the minimal load on any machine.

In this paper, we explore these classical scheduling problems within the framework of \emph{algorithms with trusted advice}. We show that limited but accurate information about an optimal solution can yield approximation guarantees beyond traditional worst-case barriers.
Specifically, in addition to problem input, the algorithm receives
a limited amount of correct information from an oracle that has access to
to a fixed optimal solution. As the
oracle reveals only partial information, the algorithm must still compute the complete schedule.
\iffalse
Much of the recent work on augmented approximation algorithms focuses on probabilistic models in which an oracle provides predictions that may be erroneous, as in learning-augmented algorithms. In contrast, we consider a deterministic setting in which the oracle provides advice that, although limited in scope, is entirely accurate.
%The oracle is accessed only once and reveals specific structural properties of an optimal solution.
This model isolates the intrinsic theoretical value of such information, avoiding the complications of error handling and allowing us to directly quantify the resulting improvements in approximation ratios.
\fi
All of our algorithms perform just a {\em single} access to the oracle. 
%This is in contrast to other learning-augmented approximation algorithms that typically perform several queries. 
The limited access to the oracle and the partial information revealed by it challenge our algorithms to ensure the information gained is impactful. 

We formalize a hierarchy of oracles, ranging from a strong oracle that reveals exact assignments of some jobs, to a weak oracle that provides minimal aggregated information. We analyze the impact of having this partial information on approximation guarantees.  

The decision regarding what information can be queried from the oracle is crucial. Known hardness results demonstrate, for instance, that revealing the actual processing times of the jobs in the {\MM} problem cannot provide a significant advantage, as they are already known in the {\sc Restricted Assignment} problem. By introducing tiny dummy jobs, it is easy to see that information regarding the exact number of jobs assigned to each machine is similarly unhelpful. Furthermore, knowing the value, $\cmaxs$, of the optimal makespan is of no use, since all currently known lower bounds remain valid. These lower bounds are all based on reductions from $3$D-matching or $3$-SAT, where knowing that a perfect matching or a satisfying assignment exists does not simplify the problem. We chose to define the oracles such that they enable improving the best-known approximation guarantees, while at the same time, as we show in Section~\ref{sec:hardness_oracles}, inapproximability bounds still hold.
%Our first main contribution addresses makespan minimization on unrelated machines. We demonstrate that by utilizing a {\em loaded machines oracle}, which simply identifies the subset of machines processing a long job in the optimal schedule, we can strictly improve upon the classical factor-$2$ barrier to achieve a $\sqrt{3}$-approximation. Furthermore, for the restricted assignment variant of the problem, we show that this oracle enables an improved $1.69$-approximation.  Our second main contribution focuses on the Max-Min Fairness problem for allocating indivisible goods. We show that a {\em satisfied players oracle}, which reveals the set of players who receive at least one highly valued item in the optimal allocation, is sufficient to guarantee a polynomial-time $2$-approximation.  Finally, we establish corresponding hardness results to map the theoretical limits of this framework. We prove that even with access to powerful oracles, such as those revealing the exact lengths of jobs or the precise optimal objective value, $R\vert{}\vert{}C_{max}$ remains NP-hard to approximate within a factor of $1.5$, and the min-max fairness problem remains NP-hard to approximate within a factor of $2$.

%\noindent{\bf Problems Statement:} 
\paragraph{Problems Studied.}

An instance of {\MM} on unrelated machines consists of a set of jobs $J = \{ 1, \ldots , n\}$ and a set of machines $M = \{1, \ldots , m \}$, along with the processing time $\pij \in {\mathbb{Z}}^+$ of each job $j$ on machine $i$, $1 \leq j \leq n$, $1 \leq i \leq m$. The goal is to assign the jobs to machines in a way that minimizes the makespan, given by the maximal total processing time assigned on one machine. Formally, an assignment is a function $\sigma:J\to M$, and the optimal makespan is
$\cmaxs=\min_{\sigma} \max_{i\in M} \sum_{ \{j \in J:\,\sigma(j)=i \}}\pij$.

The {\sc restricted assignment} problem is the special case of {\MM} where $\pij \in \{ p_j, \infty\}$, for all $1 \leq j \leq n$ and $1 \leq i \leq m$.
The special case of {\sc restricted assignment} in which every job can be assigned to at most two machines is known as {\GB}. The problem can be formulated as a graph problem as follows. Consider an undirected edge-weighted graph $G=(V,E,p)$, where $V$ represents the set of machines. Each edge $e={u,v}\in E$ corresponds to a job that can be processed only on machines $u$ and $v$, with processing time $p_e$ on either machine; its processing time on every other machine is $\infty$.
A self-loop corresponds to a job that can only be assigned to a single machine. The goal is to find an orientation of the edges that minimizes the maximum \emph{load} of all the vertices, where the load of a vertex is defined as the sum of the weights of its incoming edges. It is easy to see that the maximum load corresponds to the makespan. % of the restricted assignment problem. 

The input for the {\SC} problem is the same as the input to {\MM}: 
a set $M$ of $m$ machines, a set $J$ of $n$ jobs, and an $m \times n$ matrix of processing times. 
The goal is to assign jobs to machines in a ``fair'' way that maximizes the minimum load (total processing time) on any machine. Formally, an assignment is a function $\sigma:J\to M$, and the optimal value is $\cmins=\max_{\sigma}\min_{i\in M}\sum_{\{ j\in J:\,\sigma(j)=i \}}\pij$.
This {\em max-min fair allocation} problem is denoted {\SC} since jobs may correspond to presents, machines to kids, processing times to valuations, and the goal is to distribute presents to kids in a way that maximizes the valuation of the least satisfied kid.

%%%%%%%%%%%%%%%%%%%%%%%%%%%%%%%%%%%%%%%%%%%%%%%%%%%%%%%%%%%%%%%%%%

\subsection{Our Contributions}
\label{sec:contributions}

A main contribution of this paper is the introduction of different oracles that reveal useful information about a specific optimal solution $\OPT$. 
As detailed below, all considered problems admit improved approximation ratios even under our weakest oracle; yet, they remain NP-hard to approximate under this weakest oracle. 

For $r \geq 1$, we say that $\cA$ is an $r$-approximation algorithm for a minimization (maximization) problem $\Pi$ if, for any instance $I$ of $\Pi$, ${\cA}(I)/{\vo}\leq r$ (${\vo} /{\cA}(I) \leq r$), where ${\vo}$ is the value of an optimal solution for $I$. We first define the different oracles and then describe our results for each of them.

Given a threshold $\ell > 0$, we call a job \emph{long} if its processing time in $\OPT$ exceeds $\ell$.
We first consider the {\em long jobs schedule} oracle, $\mathsf{Oracle_{LS}}$, which reveals the assignments in $\OPT$ of all long jobs. As a simple warm up, we show that, for any $\epsilon\in(0,1]$, access to $\mathsf{Oracle_{LS}}$ yields a $(1+\epsilon)$-approximation for all of the problems considered.

Our next oracle, the {\em long-jobs representative oracle}, denoted by $\mathsf{Oracle_{LR}}$, reveals 
the assignments of a representative subset of the long jobs.
Specifically, for each machine $i \in M$ that processes at least one long job in OPT, the oracle reveals the identity of exactly one such long job. 
We show that $\mathsf{Oracle_{LR}}$ yields  approximation ratio of  $\frac{1 + \sqrt{5}}{2} \approx 1.618$ for {\MM}, and $\frac 12 + \frac 34\sqrt{2} \approx 1.561$ for {\RA}.

Our weakest oracle, the {\em long-job machines} oracle, $\mathsf{Oracle_{LM}}$, reveals the subset of machines $M_L$ that process  long jobs in $\OPT$.  This weak oracle does not indicate any assignments of jobs to these machines. We show that $\mathsf{Oracle_{LM}}$ still enables us to improve the classic bound of $2$  for {\MM}, to $\sqrt{3} \approx 1.732$. Better bounds of $\frac{\sqrt{33}+1}{4}\approx 1.686$ and $\frac{5}{8} \sqrt{7} \approx 1.6536$ are obtained for {\RA} and {\GB}, respectively. For the {\SC} problem we show a ratio of $2$.

\begin{table}[tbh]
	\centering
    \small
	\renewcommand{\arraystretch}{1.1}
	\begin{tabular}{|l|c|c|c|}
		\hline
		& $\mathsf{Oracle_{LS}}$
		& $\mathsf{Oracle_{LR}}$
		& $\mathsf{Oracle_{LM}}$ \\
		\hline
		{\MM} $(2)$
		& $\boldsymbol{1+\epsilon}$ 
		& $\boldsymbol{1.618}$ 
		& $\boldsymbol{1.732}$  \\
		\hline
		{\RA} $(2)$
		& ${\star}$
		& ${\star}$
		& $\boldsymbol{1.686}$  \\
		\hline
		{\GB} $(1.75)$
		& ${\star}$ %$(1.75)$
		& $\boldsymbol{1.561}$ 
		& $\boldsymbol{1.654}$  \\
		\hline
		{\SC} $(O(n^\epsilon))$
		& $\boldsymbol{1+\epsilon}$ 
		& ${\star}$
		& $\boldsymbol{2}$  \\
		\hline
	\end{tabular}
	\caption{\small %Our results for oracle-augmented scheduling on unrelated machines. 
    The bounds given in this paper are shown in boldface. The best-known oracle-free bounds achievable by a polynomial-time algorithm are specified in parentheses on each line. %(see also Section~\ref{sec:related}). 
    Entries marked by $\star$ follow by inference from either more general problems or weaker oracles.} %either by using upper bounds for {\MM} for its special cases, or a bound obtained for specific oracle, which carries over to a stronger oracle.}
   	\label{tab:oracle-results}
\end{table}

Table~\ref{tab:oracle-results} summarizes our results. For $\mathsf{Oracle_{LS}}$, a smaller threshold $\ell$ yields more job assignments. Thus, it is not surprising that we can approach $\OPT$ arbitrarily closely. In contrast, for the weaker oracles, choosing the parameter $\ell$ is a crucial challenge. If $\ell$ is too low, we may not be able to leverage the revealed information. The table presents the approximation ratio an oracle can yield using a single oracle access with a carefully chosen threshold.

All of our problems remain NP-hard even with the strongest oracle, $\mathsf{Oracle_{LS}}$, assuming threshold $
\ell = \Omega(\cmaxs)$. This holds since all problems are NP-hard even on instances with no jobs of length exceeding this threshold. %For our weakest oracle, $\mathsf{Oracle_{LM}}$, 
\iffalse
We show the following hardness results that match the known hardness results for oracle-free problems for our weakest oracle, $\mathsf{Oracle_{LM}}$.  
%\begin{enumerate}[wide, labelindent=0pt]
%\item[(i)]
(i) It is NP-hard to approximate {\GB} within factor better than $\frac {3} 2$, even in the presence of an oracle that reveals the value, $\cmaxs$,  of an optimal solution, and the set of vertices that receive an edge of weight more than $\frac 12 \cmaxs$. (Since {\GB} is a restricted class of {\MM}, the same bound applies also to {\MM}.)
%\item[(ii)]
(ii) It is NP-hard to approximate {\SC} within factor better than ${2}$, even in the presence of an oracle that reveals the value, $\cmins$, of an optimal solution, and the set of machines that receive a job of weight more than $\frac 12 \cmins$. Moreover, the hardness result is valid also for instances corresponding to {\sc Fair Graph Balancing}, in which every job has a fixed length and is restricted to go to one of at most two machines.
%\end{enumerate}
\fi
We prove that with our weakest oracle, $\mathsf{Oracle_{LM}}$, even the graph balancing variants of both {\MM} and {\SC} are hard to approximate. Recall that in the graph balancing variants each job can be assigned to at most two machines with identical processing time. We assume that the oracle reveals both the value of an optimal solution ($\cmaxs$ for {\MM} and $\cmins$ for {\SC}), and the set of machines that are assigned a job of length exceeding $\alpha$ times this value, for some $\alpha\in (0,1)$. We show that it is NP-hard to approximate the graph balancing variant of {\MM} within a factor better than $\frac{3}{2}$, for $\alpha \in (0,\frac 12]$, and within a factor better than $2-\alpha$, for $\alpha \in (\frac 12 ,1)$. We also show that it is NP-hard to approximate the graph balancing variant of {\SC} within a factor better than 2, for $\alpha \in (0,\frac 12]$, and within a factor better than $\frac 1\alpha$, for $\alpha \in (\frac 12 ,1)$. Our 2-approximation {\SC} algorithm consults $\mathsf{Oracle_{LM}}$ that reveals information about jobs of length at least $\frac 12$, and thus its performance matches the lower bound.

\paragraph{Organization.}
%\noindent {\bf Paper Organization:}
In the remainder of this section we highlight our techniques and overview related work.  Section~\ref{sec:makespan_unrelated} presents our results for {\MM}, and Section~\ref{sec:restricted_assign} gives the improved bounds for {\RA}. Sections~\ref{sec:graph_balance} and~\ref{sec:santa_claus} address {\GB} and {\SC}, respectively. Our hardness results appear in Section~\ref{sec:hardness_oracles}. Finally, Section~\ref{sec:discussion} concludes with a summary and directions for future work. 
%Due to space constraints, some proofs are deferred to the appendix.
%%%%%%%%%%%%%%%%%%%%%%%%%%%%%%%%%%%%%%%%%%%%%%%%%%%%%%%%%%%%%%%
\subsection{Techniques} 
\label{sec:tech}

\paragraph{Oracle-augmented makespan minimization.}
Our main technical contribution is the derivation of improved approximation 
guarantees using the long-job machines oracle, $\mathsf{Oracle_{LM}}$. We apply the technique to both {\MM} and {\RA}. Below, we outline the approach for {\MM} (see Section~\ref{sec:makespan_unrelated}).

To define which jobs are considered long in the oracle query, we choose a threshold $\ell>0$ as follows. Using the solution of the LP of~\cite{LST90} we derive a lower bound $T$ on the optimal makespan $\cmaxs$. We then introduce a parameter $\alpha\in(0,1]$, whose value is optimized later, and query $\mathsf{Oracle_{LM}}$ with threshold $\ell=\alpha T$.  

The oracle reveals only $\ml$, the set of machines that process long jobs, but does not reveal the identity of the long jobs. 
This limited information requires us to depart from the classical LP-based approximation framework. We ``guess'' that each machine in $\ml$ processes only one long job and 
formulate another LP that incorporates the information gained from $\mathsf{Oracle_{LM}}$ to obtain a stronger lower bound $\beta T$ on $\cmaxs$, where $\beta\geq 1$. 
However, this LP, which handles assignments of long as well as (possibly) short jobs to machines in $\ml$, does not enable us to obtain efficient approximation. 

We therefore consider a weaker LP in which, when formulating the
machine-load constraints, we treat the long job assigned to each machine
$i\in\ml$ as having processing time exactly $\ell$, although its actual processing time may be any value in $(\ell,\beta T]$. The resulting LP has useful structural properties: we show that any basic feasible solution can be rounded by constructing an associated bipartite graph and proving that this graph is a pseudo-forest (similar to~\cite{LST90}).%\footnote{These structural properties are used also in rounding the oracle-free LP of~\cite{LST90}.}
%However, this LP does not have the structural properties that would enable us to apply the techniques of~\cite{LST90} to round it. Consequently, 
%we then formulate a new slightly weaker LP whose constraints encode the presence of long jobs on the machines in $\ml$ and show 

Finally, through a case analysis, we derive the approximation ratio achieved by our approach as a function of the parameter $\alpha$ and optimize its value (for any value of $\beta>1$). Importantly, our analysis also accounts for an incorrect guess, namely that a machine in $M_L$ processes more than one long job.

\paragraph{Analysis for {\GB}.}
Our analysis relies on a refinement of the approach developed 
in~\cite{EKS2014} for the {\GB} problem. Using $\mathsf{Oracle_{LM}}$, we obtain 
an improved approximation guarantee by showing that the factor of $1.75$ 
established in~\cite{EKS2014} can be improved for instances that do not contain 
``excessively long'' jobs (see Section~\ref{sec:graph_balance}).

\paragraph{Oracle-augmented max-min fair allocation.} We adapt classical makespan rounding techniques for max-min fair allocation, while incorporating the information gained from $\mathsf{Oracle_{LM}}$. We formulate a customized linear program that restricts the feasible allocation space to avoid contradicting the oracle’s output while guaranteeing that machines in $M_L$ receive sufficient load from long jobs. We then round the fractional solution using a pseudo-forest bipartite graph analysis to ensure that every machine achieves a load of at least $\min\{\alpha,1-\alpha\}\cmins$, for $\alpha\in (0,1)$ (see Section~\ref{sec:santa_claus}). 

\paragraph{Hardness Results.} The classical hardness of approximation results for {\MM}~\cite{LST90} and {\GB}~\cite{AJ+11} establish a lower bound of $\frac 32$. In the instances constructed in their reductions, from $3$D-Matching and $3$-SAT, the optimal makespan is $\cmaxs=2$, jobs lengths are either 1 or 2, and length $2$ jobs are used to %block machines, 
prevent machines from receiving additional jobs. These blocked machines correspond to triplets that are not part of a perfect matching, or to literals assigned a \texttt{False} value. 
The hardness inherently collapses if an oracle reveals which machines process jobs larger than $\alpha \cmaxs$, for any $\alpha\in (0,1)$. %Modifying job sizes to $1 \pm \epsilon$ is insufficient, as $\cmaxs$ (in the reduction from $3$-SAT) may become $2+2\epsilon$, making the $(1+\epsilon)$-sized jobs not longer than $\frac 12 \cmaxs$. 
To overcome this, we introduce new reductions from $3$-SAT. Specifically, we design new variable gadgets ensuring that conflicting literals receive different assignments while maintaining $\cmaxs=2$, and guaranteeing that the machines that are assigned at least one long job in the constructed scheduling instance do not depend on the specific instance, making the oracle's output useless (see Section~\ref{sec:hardness_oracles}). %We believe these gadgets may be of independent interest for establishing further hardness results  (see Section~\ref{sec:hardness_oracles}).

\subsection{Prior Work}
\label{sec:related}

\paragraph{Oracle-free results.}
The {\MM} problem was first studied by Horowitz and Sahni~\cite{HS76}, who gave a $(1+\epsilon)$-approximation algorithm with a running time that is polynomial for a fixed number of machines but exponential in the number of machines. The first polynomial-time approximation algorithm for an arbitrary number of machines was given by Lenstra, Shmoys, and Tardos~\cite{LST90}, who achieved a factor of $2$. Shchepin and Vakhania~\cite{SV05} later improved this guarantee slightly to $2-1/m$, where $m$ is the number of machines. Nevertheless, no approximation ratio bounded away from $2$ is known. The paper~\cite{LST90} also shows that the problem is NP-hard to approximate within any factor smaller than $3/2$. Closing the gap between $3/2$ and $2$ therefore remains a longstanding open problem.

The {\RA} problem was introduced by~\cite{LST90}, whose $2$-approximation for {\MM} applies to this special case. Svensson~\cite{Svensson2012} later bounded the integrality gap of the configuration LP by $33/17+\epsilon$, although his result did not yield a polynomial-time algorithm for constructing a schedule. Jansen and
Rohwedder~\cite{JansenRohwedder2017} improved this bound to
$11/6+\epsilon$ and obtained a quasi-polynomial-time algorithm with the
same approximation ratio. Thus, no polynomial-time approximation bounded away from $2$ is known, whereas the factor-$2$ barrier can be surpassed in
quasi-polynomial time. The problem is NP-hard to approximate within any factor smaller than $3/2$~\cite{LST90}.

The {\GB} problem was introduced in~\cite{EKS2014}. The paper gives a $1.75$-approximation algorithm based on a tighter version of the LP relaxation for {\MM}~\cite{LST90}. They also show that it is NP-hard to approximate this problem with a factor better than $3/2$. A $3/2$-approximation for instances with all edge weights in 
$\{ 1,k \}$ is given in ~\cite{AJ+11}. The paper~\cite{LLP09} provides a survey of additional results, as well as an FPTAS for trees.

The {\SC} problem (max-min fair allocation) is a different well-studied objective for scheduling on unrelated machines. 
The best known polynomial-time guarantee due to Chakrabarty, Chuzhoy, and Khanna~\cite{CCK09}, is $O(n^\epsilon)$ for every fixed constant \(\epsilon >0\). 
%This result is due to  Their algorithm runs in time \(n^{O(1/\eps)}\). Thus, the guarantee can be made \(n^\eps\) for an arbitrarily small fixed \(\eps\), but no constant-factor polynomial-time approximation is known.
They show that a polylogarithmic approximation can be obtained in quasi-polynomial time (see also~\cite{BS06}). A $\sqrt{m}\log^3m$-approximation, based on a configuration LP is given in~\cite{AS10}.
%The goal is to maximizethe minimum machine completion time.
%A configuration LP can be used to obtain a $\log\log(n)/\log\log\log(n)$-approximation algorithm~\cite{BS06}. 
The paper~\cite{AFS12} presents an algorithm that approximates only the {\em value} of an optimal solution within a constant factor.  We note that the ratio of $2$ obtained in this paper with access to $\mathsf{Oracle_{LM}}$ matches the best known hardness of approximation result of $2$ for {\SC}~\cite{BD05}.
Our oracle-augmented $2$-approximation builds on an algorithm presented in~\cite{BD05}, which computes an allocation in which every machine is assigned load at most $\cmins - \max_{i,j} p_{ij}$.

\paragraph{Trusted advice.}
Trusted advice has been studied most extensively in the context of online
algorithms. In the classical advice-complexity model, an oracle that knows
the complete input sequence provides the online algorithm with a binary
advice string. The advice may describe the unknown part of the input, but
it may also encode decisions or structural properties of an optimal
offline solution. A central objective is to characterize the tradeoff
between the number of advice bits and the competitive ratio achieved by
the algorithm (see, e.g., \cite{EFKR11,BoyarEtAlSurvey}). The case of no
advice recovers the classical online model, whereas sufficiently detailed
advice can encode all the decisions of an optimal offline algorithm.

Trusted advice has received less attention in the offline setting. An
early conceptual precursor is the invited talk of Arora
\cite{Arora2000}, which proposed viewing certain approximation algorithms
as algorithms that receive a small number of correct bits about an optimal
solution from a trusted oracle. If the number of possible advice strings
is polynomial, the oracle can subsequently be eliminated by enumerating
all such strings. To our knowledge,
previous work has not systematically compared solution-based
scheduling oracles according to the structural information they disclose,
or organized such oracles into information-refinement hierarchies. Our
objective is to use such hierarchies to determine which forms of correct
structural information suffice to surpass classical approximation and
competitive-ratio barriers.

\paragraph{Learning-augmented approximation.} 
Our framework is also related to the rapidly growing literature on
learning-augmented algorithms. A learning-augmented algorithm receives, in
addition to the problem input, a possibly noisy prediction produced, for
example, from historical data or by a learned model.
Learning-augmented approximation algorithms for our problems have
only recently begun to emerge. Baba, Bampis, and Mitropoulos~\cite{BBM26} study {\MM} with predictions concerning the assignments of long jobs. Their algorithm achieves a $(1+\epsilon)$-approximation when the predictions are accurate (similar to our warmup result), while its guarantee degrades smoothly with the prediction error and never exceeds the classical factor of $2$. Xefteris~\cite{Xefteris26} considers {\RA} with a possibly infeasible predicted assignment and obtains guarantees that depend on the total processing volume that must be reassigned to reach an optimal schedule, while retaining the best oracle-free approximation guarantee as a worst-case bound. As a
special case of {\RA}, this framework applies to {\GB} as well, although it does not exploit its graph structure to obtain a problem-specific improvement. %To the best of our knowledge, no learning-augmented approximation algorithms have been developed specifically for {\GB} and for any variant of the {\SC} problem. 
To the best of our knowledge, no learning-augmented approximation algorithms exist for either {\GB} or any variant of the {\SC} problem.

\section{Minimizing Makespan on Unrelated Machines}
\label{sec:makespan_unrelated}

In this section we show improved approximation guarantees by employing the various oracles for solving the {\MM} problem.

\subsection{Warmup: The Long Jobs Schedule Oracle}
\label{sec:unrelated_strong_warmup}
Recall that for a given instance and a threshold $\ell > 0$, $\mathsf{Oracle_{LS}}$ observes a specific optimal schedule, $\OPT$, and reports  %set $J_L\subseteq J$ of jobs whose processing times in $\OPT$ is higher than $\ell$, as well as their assignment in $\OPT$. 
the assignment in $\OPT$ of all jobs with processing times greater than $\ell$ in $\OPT$.
%Let $\cmaxs$ denote the makespan of $\OPT$

\begin{theorem}
\label{thm:jlstar_strong_oracle}
$\mathsf{Oracle_{LS}}$ can be used to find a schedule with makespan at most $\cmaxs+\ell$, for any $\ell>0$.
\end{theorem}

 MOVED to APP
\begin{proof}
Given $\ell>0$, let $J_L\subseteq J$ be the set of jobs whose processing times in $\OPT$ is higher than $\ell$. $\mathsf{Oracle_{LS}}$ reports for each job in $j\in J_L$, the machine $m(j)$ to which it is assigned in $\OPT$. Consider the the LP formulation $\LPLS(t)$ in Figure~\ref{fig:LPLS} that fixes the assignment given by the oracle and is parametrized by the makespan. 
\begin{figure}[!ht]
\centering
\fbox{%
\begin{minipage}{\dimexpr 0.75\linewidth - 2\fboxsep - 2\fboxrule \relax}
{\small
\begin{align}
\LPLS(t) 
\quad & \displaystyle\sum_{i \in M} \xij =1  &&\quad \forall j \in J \label{rev:constraint_jobs_LS} \\
\quad & \displaystyle\sum_{j \in J} \pij \xij \leq t, &&\quad \forall i \in M \label{rev:machine_comptime_LS} \\
\quad & \xij = 1  &&\quad \forall  j  \in J_L\, \wedge\, i=m(j) \nonumber \\
\quad & \xij = 0  &&\quad \forall  j  \in J_L\, \wedge\, i\ne m(j) \nonumber \\ 
\quad & \xij = 0  &&\quad \forall  j  \in J\setminus J_L\, \wedge\, i\in M \text{ for which } \pij>\ell\nonumber \\
\quad & \xij \geq 0  &&\quad \forall j  \in J\setminus J_L \nonumber 
\end{align}
}
\end{minipage}
}
\caption{The LP formulation $\LPLS(t)$}
\label{fig:LPLS}
\end{figure}

Find the minimum integer $T$ such that $\LPLS (T)$ has a feasible solution. This can be done in polynomial time by binary search. Note that $T\le\cmaxs$. Find a basic solution of  $\LPLS(T)$. Round this solution using the technique of \cite{LST90}. Since for any fractional $\xij$ in this solution, the length of job $j$ must be at most $\ell$, the resulting schedule has makespan at most $T + \ell\le\cmaxs +\ell$. 
\end{proof}

The induced approximation ratio is $1+\frac{\ell}{\cmaxs}$. Thus, to achieve a $(1+\epsilon)$-approximation, we can query $\mathsf{Oracle_{LS}}$ with $\ell \le \epsilon T$ 
where $T\le \cmaxs$ is the minimum integer for which $\LPLS (T)$ has a feasible solution.

\subsection{The Long Jobs Representative Oracle}
\label{sec:unrelated_warmup1}
We consider the oracle $\mathsf{Oracle_{LR}}$ that for a given instance and a threshold $\ell > 0$, observes a specific optimal schedule, $\OPT$, and reveals, for each machine $i \in M$ that processes at least one  job with processing time greater than $\ell$, the identity of exactly one such job. We call the jobs whose assignments are revealed by $\mathsf{Oracle_{LR}}$ \emph{representative} jobs, and denote the set of these jobs by $J_R\subseteq J$. For each $j\in J_R$, let $m(j)$ denote the machine assigned to it in $\OPT$.
%set $J_L\subseteq J$ of jobs whose processing times in $\OPT$ is higher than $\ell$, as well as their assignment in $\OPT$. 
%the assignment in $\OPT$ of all jobs with processing times higher than $\ell$ in $\OPT$.the {\em partial assignment} oracle accepts, as $\mathsf{Oracle_{LS}}$, a threshold $\ell>0$, and returns as before $J_L$. In addition, For each machine $i \in M$ which processes at least one job in $J_L$, the oracle indicates the index of {\em one} such job.

%(((((((( OLD - with $\alpha$ ))))$\alpha \in (0,1]$ and $T \geq \pmax$, and defines as before $J_L$. For each machine $i \in M$ which processes at least one job in $J_L$, the oracle indicates the index of {\em one} such job. )))
  
\begin{theorem}
\label{thm:strong_oracle}
$\mathsf{Oracle_{LR}}$ yields a golden ratio $\varphi\approx 1.618$-approximation to the minimum makespan.
\end{theorem}

\begin{proof}
Recall the LP formulation of~\cite{LST90} given in Figure~\ref{fig:lplst}. %, where all machines are available for $t$ time units.
\begin{figure}[!ht]
\centering
\fbox{%
\begin{minipage}{\dimexpr 0.75\linewidth - 2\fboxsep - 2\fboxrule \relax}
{\small
\begin{align}
\textbf{LPLST}(t) 
\quad & \displaystyle\sum_{i \in M_j[t]} \xij =1  &&\quad \text{for } j = 1, \ldots, n \label{rev:constraint_jobs} \\
\quad & \displaystyle\sum_{j \in J_i[t]} \pij \xij \leq t &&\quad \text{ for } i = 1, \ldots, m \label{rev:machine_comptime} \\
\quad & \xij \geq 0  &&\quad \text{for } j  \in J_i(t), i = 1, \ldots, m \nonumber 
\end{align}
}
\end{minipage}
}
\caption{The LP formulation $\textbf{LPLST}(t)$}
\label{fig:lplst}
\end{figure}

In this LP, the notation $J_i[t]$, $t \in \intpos$, denotes the subset of jobs with processing time less than or equal to $t$ on machine $i$, and the notation $M_j[t]$, $t \in \intpos$, denotes the subset of machines on which the processing time of job $j$ is less than or equal to $t$. 
%Similarly, let $J_i(s,t)$, $s,t \in \intpos $, denote the subset of jobs with processing times at least $s$ and at most $t$ on machine $i$. 
Lenstra et al.~\cite{LST90} find the minimum integer, denoted $T$, for which \textbf{LPLST}$(t)$ is feasible, and output a schedule with makespan $2T$.

%For some $\alpha \in (1/2,1)$ to be fixed later, we say that job $j$ is {\em long} on machine $i$ if $p_{ij} \ge \alpha T$; else, job $j$ is {\em short}. 
Given an instance of {\MM}, we apply~\cite{LST90} to find $T$ defined above.
We query $\mathsf{Oracle_{LR}}$ with threshold $\ell=\alpha\cdot T$, for some $0<\alpha<1$, to be fixed later.

We ``guess'' that the set $J_R$ revealed by $\mathsf{Oracle_{LR}}$ is the set of \emph{all} jobs with processing time greater than $\ell$ in $\OPT$. 
Assuming this guess is true we go ahead and use the LP formulation $\LPLS(t)$ defined above (where $J_R$ plays the role of $J_L$) to try to find the minimum integer $T'$ such that $\LPLS (T')$ has a feasible solution. If no such $T'$ exists then we have an evidence that our guess is wrong. If such a $T'$ exists, then as described above, we obtain a schedule with makespan at most $T' + \ell$. Note that $T'\ge T$ since any feasible solution of $\LPLS(t)$ is also a feasible solution of $\textbf{LPLST}(t)$. Moreover, if our guess is correct we are also guaranteed that $T'$ is at most the makespan of $\OPT$, which implies an approximation ratio of $\frac{T'+\alpha T}{T'}\le 1+\alpha$.

If our guess is wrong, then it must be the case that at least one machine in $M_R$ processes at least \emph{two} jobs with processing time greater than $\ell$. This implies that the makespan of $\OPT$ is at least $2\ell$. Since we already obtained a valid schedule of makespan $2T$ using~\cite{LST90}, we get that in this case the approximation ratio is $\frac{2 T}{2\alpha T}= \frac 1\alpha$.

We output the best schedule among the schedule with makespan $T' + \ell$ (if such exists) and the schedule with makespan $2T$. Since we cannot control whether our guess is correct or not, the resulting approximation ratio of this schedule is $\max\{1+\alpha, \frac 1\alpha\}$. The optimal $\alpha$ is achieved $1+\alpha= \frac 1\alpha$, implying a ratio $1+\alpha = \varphi= \frac{1 + \sqrt{5}}{2} \approx 1.618$
%
% Let $\varphi= \frac{1 + \sqrt{5}}{2} \approx1.618$ be the golden ratio,
%  and $T=d_1=d_2= \cdot \cdot \cdot = d_m$ be a minimum uniform upper bound on the completion times of all machines; that is, $LP(T)$ has a feasible solution, but 
%   $LP(T-1)$ has no feasible solution. We access $\mathsf{Oracle_{LR}}$ with parameter $\ell = T/\varphi$. For any machine that processes a job in $J_L$, the oracle reveals exactly one of these jobs.
% We distinguish between two cases.
% \begin{enumerate}
% \item 
% Each machine processes at most one long job. We modify the last set of constraints in $LP(t)$ by setting $\xij =1$ for each job $j \in J_L$ and the machine $i \in M$  on which this job is processed. We now find the minimum value $T'$ such that the modified program ${LP}_{\cal M} (T')$ has a feasible solution, but ${LP}_{\cal M} (T' -1)$ has no feasible solution. We note that as the modified constraints in ${LP}_{\cal M} (T')$ are derived from an integral solution, it holds that $T' \geq T$. Solving ${LP}_{\cal M} (T')$ and rounding using the technique of \cite{LST90}, we have a makespen at most $T' + \frac{T}{\varphi}\leq T'(1 + 1/\varphi) = \varphi T' \leq \varphi \cmaxs$, where $\cmaxs$ is the optimal makespan.
% \item
% At least one machine processes more than one long job. Then the optimal makespan satisfies: $\cmaxs \geq \frac{2T}{\varphi}$. Solving $LP(t)$, we have a makespan at most $2T$, which yields the desired ratio.
% \end{enumerate}
%
\end{proof}

\subsection{The Long Job Machines Oracle}
Next, we consider our weakest oracle $\mathsf{Oracle_{LM}}$ that does not provide specific job-to-machine assignments and show that it still improves the approximation ratio of $2$ achieved in~\cite{LST90}. For a given instance and a threshold $\ell>0$, the oracle $\mathsf{Oracle_{LM}}$ observes a specific optimal
schedule, $\OPT$, and reveals the subset of machines, denoted $M_L$, such that each machine $i\in M_L$ processes at least one job with processing time greater than $\ell$ in $\OPT$. 

Given an instance of {\MM}, we again apply~\cite{LST90} to find the minimum integer, $T$, for which \textbf{LPLST}$(t)$ is feasible.
We query $\mathsf{Oracle_{LM}}$ with threshold $\ell=\alpha\cdot T$, for some $0<\alpha<1$, to be fixed later, and are given the respective set $M_L$. 
%We say that job $j$ is {\em long} on machine $i$ if $p_{ij} \ge \ell$; else, job $j$ is {\em short} on machine $i$. Note that if $i\notin M_L$ then machine $i$ is not assigned any long job on this machine.

%Given an optimal schedule, $\OPT$, the oracle indicates $M_L$, the subset of machines on which a long job is assigned in $\OPT$. 
%To obtain an improved approximation ratio we handle separately the case where some machine in $M_L$ is assigned more than one long job on it
Similar to Section~\ref{sec:unrelated_warmup1}, we ``guess'' that each machine in $M_L$ is assigned {\em exactly} one job with processing time greater than $\ell$.
%As before, let $T \in \intpos$ be the minimal value for which LP(T) has a feasible solution. For a given $\alpha \in [1/2, 1]$, we say that job $j$ is {\em long} on machine $i$ if $\pij \in [\alpha T,t]$, where $t \in \intpos$ is a parameter. Given an optimal schedule, $\OPT$, the oracle indicates $M_L$, the subset of machines on which a long job is assigned in $\OPT$. We seek a schedule of all the jobs such that each machine in $M_L$ is assigned exactly one long job, and the completion time of machine $i$ is at most $d_i$, for all $ 1 \leq i \leq m$. We denote the input for our scheduling problem by ${\cI} = (J,M,M_L,P, {\bard},{\bartau})$. 
%
%We now define the LP relaxation of this problem. To this end, recall the LP formulation of~\cite{LST90} (where $d_i=t$).
%
%\begin{align}
%\textbf{LP}(t) 
%\quad & \displaystyle\sum_{i \in M_j(t)} \xij =1  \quad \text{for } j = 1, \ldots, n \label{rev:constraint_jobs} \\
%\quad & \displaystyle\sum_{j \in J_i(t)} \pij \xij \leq t, \text{ for } i = 1, \ldots, m \label{rev:machine_comptime} \\
%\quad & \xij \geq 0  \quad \text{for } j  \in J_i(t), i = 1, \ldots, m \nonumber 
%\end{align}
%In the above formulation, the notation $J_i(t)$, $t \in \intpos $, denotes the subset of jobs with processing times at most $t$ on machine $i$. Similarly, let $J_i(s,t)$, $s,t \in \intpos $, denote the subset of jobs with processing times at least $s$ and at most $t$ on machine $i$. Let $T$ be the minimum integer for which \textbf{LP}$(t)$ is feasible. 
Assuming this guess is true we
define a new LP formulation $\wtlp(t)$ %that uses $T$ and a constant $\alpha \in (\frac 12,1)$, to be fixed later, 
given in Figure~\ref{fig:wtlp}.

\begin{figure}[htbp]
\centering
\fbox{%
\begin{minipage}{\dimexpr 0.75\linewidth - 2\fboxsep - 2\fboxrule \relax}
{\small
\begin{align}
\wtlp(t)
\quad & \displaystyle\sum_{i \in M_j[t]} \xij =1  &&\quad \text{for } j=1, \ldots , n  \label{rev:full_assign}\\ 
\quad & \displaystyle\sum_{j \in J_i[\alpha T,t]} \xij =1 &&\quad  \text{ for } i \in M_L \label{rev:matching}\\
\quad & \displaystyle\sum_{j \in J_i(\alpha T)} \pij \xij \leq t &&\quad \text{ for } i \in M \setminus M_L  \label{rev:wtlp_bml} \\ 
\quad & \displaystyle\sum_{j \in J_i[t-\alpha T] \cup J_i[\alpha T,t]} \pij \xij \leq t &&\quad \text{ for }   i \in M_L  \label{rev:wtlp_ml} \\ 
\quad & \xij = 0  &&\quad \text{    for } i \in M \setminus M_L ,\, j \in J_i[\alpha T,t] \label{rev:bmlzero} \\
\quad & \xij = 0  &&\quad \text{    for } i \in M_L ,\, j \in J_i(t-\alpha T,\alpha T) \label{rev:mlzero} \\
\quad & \xij \geq 0  &&\quad \text{    for } i \in M ,\, j \in J_i[t] \label{rev:non_negative_all}
\end{align}
}
\end{minipage}
}
\caption{The LP formulation $\wtlp(t)$}
\label{fig:wtlp}
\end{figure}

In $\wtlp(t)$ the notation $J_i[s,t]$, $s,t \in \intpos $, denotes the subset of jobs with processing times in the (closed) interval $[s,t]$ on machine $i$. 
Similarly, the notation $J_i(s,t)$, $s,t \in \intpos $, denotes the subset of jobs with processing times in the (open) interval $(s,t)$ on machine $i$, and the notation $J_i(t)$, $t \in \intpos $, denotes the subset of jobs with processing times in the (open) interval $(0,t)$ on machine $i$. 

Constraint~(\ref{rev:full_assign}) ensures that every job is assigned on some machine on which its length is at most $t$. 
Constraint~(\ref{rev:matching}) restricts every machine in $M_L$ to processes exactly one long job.
Constraint~(\ref{rev:wtlp_bml}) ensures that every machine that does not process a long job processes short jobs of total length at most $t$. 
Constraint~(\ref{rev:wtlp_ml}) bounds to $t$ the total load on every machine that processes a long job. Since a long job has length at least $\alpha T$, only jobs whose length is at most $t-\alpha T$ can be processed in addition to a long job.
Constraint~(\ref{rev:bmlzero}) ensures that every machine in $M\setminus M_L$ is not assigned a job with long processing time on this machine.
Constraint~(\ref{rev:mlzero}) ensures that every machine in $M\setminus M_L$ is not assigned a job with processing time $t \in (t-\alpha T,\alpha T)$ on this machine. (We remark that this constraint is vacuous if $t\geq 2\alpha T$.)
Finally, Constraint~(\ref{rev:non_negative_all}) relaxes the integrality of the binary variables $\xij$. 

We now find the minimum integer $T'$ for which $\wtlp(t)$ is feasible. Note that $\wtlp$ may not be feasible for any value $t$. If there is a feasible solution then it is easy to see that $T' \ge T$ since every feasible solution of $\wtlp(t)$ is also a feasible solution of $\mathbf{LPLST}(t)$. To see this note that constraints (\ref{rev:wtlp_ml}) and (\ref{rev:wtlp_bml}) together with constraints (\ref{rev:bmlzero}) and (\ref{rev:mlzero}) are tighter than constraint~(\ref{rev:machine_comptime}) since the summations in constraints~(\ref{rev:wtlp_ml}) and~(\ref{rev:wtlp_bml}) are done on subsets of the respective sets in the summation in constraint (\ref{rev:machine_comptime}), and the variables that are not in the subsets defined in constraints~(\ref{rev:wtlp_ml}) and~(\ref{rev:wtlp_bml}) are fixed to zero. 

%Proofs of the following is identical to what we had before. \textcolor{blue}{[The lemma and corollary hold for any parameter $t > 0$.]} 
%\begin{lemma}
%	\label{lem:extreme_pos_vars1}
%	Any extreme point of $\wtlp(t)$ has at most $n+m+|M_L|$ entries  with strictly positive values.
%\end{lemma}
%\begin{corollary}
%\label{cor:extreme_frac_assigned1}
%In any extreme point of $\wtlp(t)$ at most $m + |M_L|$ jobs are fractionally assigned.
%\end{corollary}

%We now prove the approximation ratio.
\begin{theorem}
	\label{thm:loaded_machines_oracle1}
	$\mathsf{Oracle_{LM}}$ yields a $\sqrt 3$-approximation to the minimum makespan. %, for any $m  > 1$.
\end{theorem}
\begin{proof}
Recall that $T'$ is the minimum integer value for which $\wtlp(T')$ has a feasible solution.
%Let OPT be an optimal solution in which the set of machines that are assigned long jobs is exactly $M_L$.
We distinguish between two cases.
\par\noindent\textbf{Case 1:}
${\wtlp}(t)$ has no feasible solution. This implies that our guess is false, and at least one machine $i\in M_L$ is assigned more than one long job in $\OPT$.  It follows that $\cmaxs > 2 \alpha T$. Since the schedule given by rounding $\textbf{LPLST}$ as in \cite{LST90} has makespan at most $2T$, we have an approximation ratio of at most $\frac 1\alpha$. 
\par\noindent\textbf{Case 2:}
${\wtlp}(t)$ has a feasible solution, and $T'$ is the minimum integer for which $\wtlp(t)$ is feasible. In this case $T'$ is a lower bound on the makespan of any schedule in which every machine in $M_L$ is assigned exactly one long job on it. We distinguish between two sub-cases based on the value of $T'$.
\par\noindent\textbf{Case 2.1:}
$T' \geq  2\alpha T$. In this case $2 \alpha T$ is a lower bound on the makespan of any feasible schedule since it also lower bounds the makespan of any schedule in which at least one machine in $M_L$ is assigned more than one long job and we have an approximation ratio of at most $\frac 1\alpha$. %(Note that in this case in the constraints within (\ref{rev:wtlp_ml}) some jobs may be double counted on the left hand side, however this does not change the fact that $\cmaxs > 2 \alpha T$.)
\par\noindent\textbf{Case 2.2:}
$T'< 2\alpha T$. 
In this case we formulate another LP that is slightly weaker than $\wtlp$ but can be rounded following the techniques of~\cite{LST90}.
Define a new linear program denoted $\hatlplm$, given in Figure~\ref{fig:hatlplm}, that uses $T$, $T'$ and $\alpha$.

\begin{figure}[htbp]
\centering
\fbox{%
\begin{minipage}{\dimexpr 0.75\linewidth - 2\fboxsep - 2\fboxrule \relax}
{\small
\begin{align}
\hatlplm 
\quad & \displaystyle\sum_{i \in M_j[T']} \xij =1  &&\quad \text{for } j=1, \ldots , n  \label{hat:full_assign}\\ 
\quad & \displaystyle\sum_{j \in J_i[\alpha T,T']} \xij =1 &&\quad  \text{ for } i \in M_L \label{hat:matching}\\
\quad & \displaystyle\sum_{j \in J_i(\alpha T)} \pij \xij \leq T' &&\quad \text{ for } i \in \bml  \label{hat:hatlp_bml} \\ 
\quad & \displaystyle\sum_{j \in J_i[T'-\alpha T]} \pij \xij \leq T'-\alpha T &&\quad \text{ for }   i \in M_L  \label{hat:hatlp_ml} \\ 
\quad & \xij = 0  &&\quad \text{    for } i \in M \setminus M_L ,\, j \in J_i[\alpha T,T'] \label{rev:hatbmlzero} \\
\quad & \xij = 0  &&\quad \text{    for } i \in M_L ,\, j \in J_i(T'-\alpha T,\alpha T) \label{rev:hatmlzero} \\
\quad & \xij \geq 0  &&\quad \text{    for } i \in M , j \in J_i[T']
\label{hat:non_negative_all}
\end{align}
}
\end{minipage}
}
\caption{The LP formulation $\hatlplm$}
\label{fig:hatlplm}
\end{figure}

Constraint~(\ref{hat:full_assign}) ensures that every job is assigned on some machine on which its length is at most $T'$. 
Constraint~(\ref{hat:matching}) determines that every machine in $M_L$ processes exactly one long job.
Constraint~(\ref{hat:hatlp_bml}) ensures that every machine that does not process a long job, processes short jobs of total length at most $T'$. 
Constraint~(\ref{hat:hatlp_ml}) bounds by $T'-\alpha T$ the contribution of short jobs to the load on every machine that processes a long job.
Constraint~(\ref{rev:hatbmlzero}) ensures that every machine in $M\setminus M_L$ is not assigned a job with long processing time on this machine.
Constraint~(\ref{rev:hatmlzero}) ensures that every machine in $M\setminus M_L$ is not assigned a job with processing time $t \in (T'-\alpha T,\alpha T)$ on this machine. (Note that $T'-\alpha T<\alpha T$ since $T'< 2\alpha T$.)
Finally, Constraint~(\ref{hat:non_negative_all}) relaxes the integrality of the binary variables $\xij$.

Note that any feasible solution of $\wtlp(T')$ is also a feasible solution of $\hatlplm$ (but not vice versa) since constraint~(\ref{hat:hatlp_ml}) is weaker than constraint~(\ref{rev:wtlp_ml}). Since constraints~(\ref{rev:hatbmlzero}) and~(\ref{rev:hatmlzero}) of $\hatlplm$ fix some variables to zero, we omit these variables from the formulation and disregard these constraints. Consequently and since $T'-\alpha T<\alpha T$, every variable of $\hatlplm$ appears in exactly one constraint within constraints~(\ref{hat:matching}), (\ref{hat:hatlp_bml}), and~(\ref{hat:hatlp_ml}). We use this fact in the analysis below.

The following results will play a central role in the analysis of the rounding subroutine applied by our algorithm.
\begin{claim}
	\label{claim:extreme_pos_vars1}
	Any extreme point of $\hatlplm$ has at most $n+m+|M_L|$ entries  with strictly positive values.
\end{claim}

\begin{proof}
	In $\hatlplm$ there are $r=O(mn)$ variables and five types of constraints: $n$ constraints of type (\ref{hat:full_assign}), $|M_L|$ constraints of type
	(\ref{hat:matching}), $|\bml|$ constraints of type (\ref{hat:hatlp_bml}), $|M_L|$ constraints of type (\ref{hat:hatlp_ml}), and $r$ constraints of type (\ref{hat:non_negative_all}). Let $\bx$ be an extreme point of $\hatlplm$. Then there are $r$ linearly independent constraints which are tight in $\bx$. Out of these constraints, there are at most $(n+m+|M_L|)$ constraints of types 
	(\ref{hat:full_assign}), (\ref{hat:matching}), (\ref{hat:hatlp_bml}), and (\ref{hat:hatlp_ml}).
	Therefore, there are at least $(r-n-m-|M_L|)$ constraints of type (\ref{hat:non_negative_all}) which are tight, i.e., $(r-n-m-|M_L|)$ entries in $\bx$ that are equal to zero. Thus, at most $n+m+|M_L|$ entries are strictly positive. 
\end{proof}
\begin{corollary}
	\label{cor:extreme_frac_assigned1}
	In any extreme point of $\hatlplm$ at most $m + |M_L|$ jobs are fractionally assigned.
\end{corollary}

\begin{proof}
	Let $\bx$ be an extreme point of $\hatlplm$, and let $k$ denote the number of strictly positive entries in $\bx$. By Claim~\ref{claim:extreme_pos_vars1}, $k \leq n+m+|M_L|$. We note that $k$ corresponds to the number of assignments of jobs to machines in $\bx$. Each job requires at least one assignment, and a job $j$ needs more than one assignment if $j$ is assigned fractionally. Thus, the number of jobs that are assigned fractionally in $\bx$ is at most $k-n \leq m+|M_L|$.
\end{proof}

Let $\bx$ be a basic feasible solution for $\hatlplm$. 
%Note that for each machine $i \in M$ and job $j \in J$, if the variable $\xij$ is defined, it appears only in a single constraint within (\ref{hat:hatlp_bml}),(\ref{hat:hatlp_ml}), and  (\ref{hat:matching}). We note that since we are in Case 2.2, it holds that $T' - \alpha T < \alpha T$. Hence, for every machine $i \in M_L$, the sets $J_i[T' - \alpha T]$ and $J_i[\alpha T,T']$ are disjoint. It follows that each variable $x_{ij}$ appears in exactly one of the constraints (\ref{hat:hatlp_bml}), (\ref{hat:hatlp_ml}), and (\ref{hat:matching}). Therefore, 
Recall that each variable $x_{ij}$ appears in exactly two constraints of $\hatlplm$:
one assignment constraint (within~(\ref{hat:full_assign})) and one machine-side constraint (within~(\ref{hat:matching}), (\ref{hat:hatlp_bml}), and~(\ref{hat:hatlp_ml})). 
We leverage on this property to construct the following bipartite graph $G=(J,C,E_G)$. Each vertex $j \in J$ corresponds to a job. The set $C$ corresponds to the constraints within~(\ref{hat:matching}), (\ref{hat:hatlp_bml}), and~(\ref{hat:hatlp_ml}), and contains a single vertex for each machine $i \in \bml$ and two vertices for each machine $i \in M_L$, denoted $i$ and $\hati$. Each edge in $E_G$ corresponds to a variable $\xij$ that is strictly positive in $\bx$. 
If $i \in \bml$ and job $j \in J_i(\alpha T)$ then we add an edge $(j,i) \in E_G$. Similarly, if $i \in M_L$ and job $j \in J_i[T'- \alpha T]$ then we add an edge $(j,i) \in E_G$. Lastly, if $i \in M_L$ and job $j \in J_i[\alpha T,T']$ then we add an edge $(j,\hati) \in E_G$. 
The overall number of edges in $G$ is equal to the size of the support of $\bx$, i.e., the number of strictly positive variables in $\bx$. It follows that $G$ has $(n +m + |M_L|)$ vertices, and by Claim~\ref{claim:extreme_pos_vars1}, $|E_G| \leq n+m+ |M_L|$. 

A graph is a {\em pseudo-tree} if it is connected and contains at most one cycle, which implies that it is either a tree, or a tree plus an extra edge. A graph is a {\em pseudo-forest} if each of its connected components is a pseudo-tree. 
\begin{claim}
\label{cl:pseudo_forest}
The bipartite graph $G=(J,C,E_G)$ is a pseudo-forest.
\end{claim}
\begin{proof}
We distinguish between two cases. 
%\begin{itemize}[wide, labelindent=0pt]
\par\noindent\textbf{Case 1:}
$G$ is connected. Then $G$ must contain a subgraph that is a spanning tree. The spanning tree has $n+m+ |M_L|-1$ edges. Hence, $G$ is a spanning tree plus at most one extra edge, i.e., a pseudo-tree.
\par\noindent\textbf{Case 2:}
$G$ is not connected. We show that in this case every connected component of $G$ is a pseudo-tree. Assume towards a contradiction that there is a connected component  $H=(J_H, C_H, E_H)$ that is not a pseudo-tree, where $J_H \subseteq J$, $C_H \subseteq C$, and $E_H \subseteq E_G$. Denote by $\bh= (J_{\bh},C_{\bh},E_{\bh})$ the rest of the graph $G$; that is,
$J_{\bh}=J \setminus J_H$, $C_{\bh}=C \setminus C_H$, and  $E_{\bh}=E_G \setminus E_H$.

We now use $H$ and $\bh$ to define two restricted LPs. The first LP,
denoted $\hatlp_H$, consists of all variables $\xij$ whose corresponding edge belongs to $E_H$. The constraints of $\hatlp_H$ are derived by restricting the constraints of $\hatlplm$ to the variables of $\hatlp_H$. Specifically, the constraints of $\hatlp_H$ are the constraints of $\hatlplm$ within~(\ref{hat:full_assign}) indexed by the jobs in $J_H$ restricted to the variables whose corresponding edge belongs to $E_H$, together with the constraints indexed by $C_H$ within~(\ref{hat:matching}), (\ref{hat:hatlp_bml}), and~(\ref{hat:hatlp_ml}) restricted to the variables whose corresponding edge belongs to $E_H$.  %together with all constraints of $\hatlplm$ indexed by the vertices in $J_H\cup C_H$. 
The second LP, denoted $\hatlp_{\bh}$, is defined similarly with respect to all variables $\xij$ whose corresponding edge belongs to $E_{\bh}$. Specifically, the constraints of $\hatlp_{\bh}$ are the constraints of $\hatlplm$ within~(\ref{hat:full_assign}) indexed by the jobs in $J_{\bh}$ restricted to the variables whose corresponding edge belongs to $E_{\bh}$, together with the constraints indexed by $C_{\bh}$ within~(\ref{hat:matching}), (\ref{hat:hatlp_bml}), and~(\ref{hat:hatlp_ml}) restricted to the variables whose corresponding edge belongs to $E_{\bh}$. 
%together with all constraints of $\hatlplm$ indexed by the vertices in $J_{\bh}\cup C_{\bh}$.
		
We claim that the set of strictly positive entries of $\bx$ can be split into
two vectors, $\bxh$ and $\bx_{\bh}$, which are feasible solutions for
$\hatlp_H$ and $\hatlp_{\bh}$, respectively. Indeed, for any positive entry
$\xij$, the corresponding edge belongs either to $E_H$ or to $E_{\bh}$.
Moreover, because $H$ and $\bh$ are different connected components, the two
constraints in which $\xij$ participates are both indexed by vertices of the
same component. Hence, the variables corresponding to $E_H$ appear only in the
constraints indexed by $J_H\cup C_H$, and the variables corresponding to
$E_{\bh}$ appear only in the constraints indexed by
$J_{\bh}\cup C_{\bh}$. Since $\bx$ is feasible for $\hatlp$, it follows
that $\bxh$ is feasible for $\hatlp_H$ and $\bx_{\bh}$ is feasible for
$\hatlp_{\bh}$.
		
Next, we show that $\bxh$ cannot be an extreme point of $\hatlp_H$. Since $H$
is connected but not a pseudo-tree, it contains at least two cycles. Therefore,
$|E_H| > |J_H|+|C_H|$.
Now consider the constraint matrix of $\hatlp_H$. We note that each variable $\xij$ is associated with a column in the constraint matrix. By the definition of $H$, the columns of the constraint matrix corresponding to the variables in $E_H$ have nonzero entries only in the rows indexed by $J_H\cup C_H$. Therefore, these columns lie in a space of dimension at most $|J_H|+|C_H|$. Since the number of columns associated with $\hatlp_H$ satisfies $|E_H|>|J_H|+|C_H|$, these columns
are linearly dependent. Recall that any basic feasible solution for a linear program corresponds to a set of linearly independent columns. Hence, $\bxh$ is not a basic solution of $\hatlp_H$.

Thus, $\bxh$ can be written as a convex combination of other feasible points; that is, $\bxh = \lambda \byo + (1- \lambda) \byt$, where $\byo \neq \byt$. Then $\bx$ can also be written as the convex combination of the feasible points of the original instance, i.e., $\bx = \lambda (\bxhath + \byo) + (1- \lambda)(\bxhath + \byt)$. A contradiction, since $\bx$ is an extreme point. Therefore, $G$ is a pseudo-forest. 
\end{proof}

We now show how to obtain a feasible schedule by rounding the solution $\bx$. First, for each variable $\xij=1$ in $\bx$
we assign job $j$ to machine $i$. Also, we remove the vertex $j$ from $J$ and the single edge touching $j$ from $E_G$. If the resulting subgraph has isolated vertices in $C$ we remove them as well.

The resulting subgraph $G'= (J',C', E_G')$ is still a pseudo forest. Also, each vertex in $J'$ has degree at least two. This is because each edge corresponds to a variable $\xij$ that is strictly less than 1 while each job needs to be fully assigned. 

We assign each of the jobs that correspond to vertices in $J'$ to machines by finding a matching that matches all the vertices in $J'$ to vertices in $C'$ and assigning a job $j$ to the machine matched to $j\in J'$. It follows that each of these jobs is assigned to a different machine.

The matching is found as follows. Suppose that there exists a vertex $i\in C'$ with degree 1, and let $(j,i)$ be the unique edge in $E_G'$ that touches vertex $i$. We  match  $j$ to  $i$. Also, we remove the vertex $j$ from $J'$ and all the edges touching $j$ from $E_G'$. We also remove $i$ and any other isolated vertex in $C'$ from the resulting subgraph. We continue this process until there are no vertices of degree 1 in $C'$.

At this point all the vertices in $C'$ are of degree higher than 1. Since all the vertices in $J'$ are also of degree higher than 1 and $G'$ is a pseudo forest, $G'$ must be a collection of disjoint (even) cycles, and thus $G'$ has a perfect matching.

All left is to analyze the quality of the resulting rounded solution. Note that every machine $i\in \bml$ is assigned at most one job for which $\xij<1$. It follows that the makespan of machine $i$ is bounded by $T'+\alpha T$. Similarly, every machine $i\in M_L$ is assigned at most one job $j$ that satisfies (i) the processing time of job $j$ on  machine $i$ is at most $T'-\alpha T$, and (ii) $\xij<1$. It follows that the total processing time on machine $i$ of all jobs assigned to this machine with processing time at most $T'-\alpha T$ is bounded by $2 (T'-\alpha T)$.

We still need to handle the long jobs on $M_L$. Consider a machine $i\in M_L$. We claim that machine $i$ is assigned at most one job from $J_i[\alpha T,T']$. Suppose that $j$ is such a job. 
%Recall that in this case $\xij$ appears in a single constraint within (\ref{hat:matching}). 
If $\xij=1$ then after assigning job $j$ to $i$, the vertex in $C$ that corresponds to this constraint becomes isolated and removed from $G'$ and thus no other long job is assigned to machine $i$. If $\xij<1$ then there is no long job $j'$ such that $x_{i,j'}=1$ (since in this case $i$ would not be in $C'$). Since each of the jobs in $J'$ is matched to a different vertex in $C'$, no other job $j'\in J_i[\alpha T,T']$, such that $x_{i,j'}<1$ is assigned to machine $i$. Recall that the processing time of job $j$ on machine $i$ is at most $T'$. 
By combining this with the analysis above, we get that the makespan of machine $i$ is bounded by $2 (T'-\alpha T) + T' = 3 T'-2\alpha T$.

To analyze the approximation ratio in Cases 2.1 and 2.2, note that the makespan of the optimal solution chosen from schedules in which exactly one long job is assigned to each machine in $M_L$ is at least $T'$. By taking the minimum between the schedule given by rounding the original $\textbf{LPLST}(T)$ and the schedule given by rounding a basic solution for $\hatlplm$, the makespan of the approximate solution is $\min\{2T,\max\{T'+\alpha T,3 T'-2\alpha T\}\}$, where $T'\ge T$. Substitute $T'=\beta T$, for some $\beta>1$. Combining the bounds from Case 1 and Cases 2.1 and 2.2 we get that the approximation ratio is bounded by 
\[
R= \max \{\frac{1}{\alpha}, 
    \min \{\frac{2}{\beta},
        \max \{1+\frac\alpha\beta,3-2\frac\alpha\beta\}\}\},
\]   
for any $\beta\ge 1$, while we can optimize $\alpha\in (0,1)$. 
By standard analysis, the optimal value is $\alpha = \sqrt{3}-1$, giving $R\le \sqrt{3}$.
\iffalse
Fix $\alpha = \sqrt{3}-1$. For this value $\frac{1}{\alpha} \approx 1.366 < \sqrt{3}$. We show that the maximum of the function 
$f(\beta) =\min \{\frac{2}{\beta},
\max \{1+\frac\alpha\beta,3-2\frac\alpha\beta\}\}$, for this fixed value of $\alpha$ is $\sqrt{3}\approx 1.732$. The value of $f(1)$ is determined by $1+(\sqrt{3}-1)=\sqrt{3}$.
The value of $f(\cdot)$ in the interval $(1,2/\sqrt{3})$ is upper bounded by $\sqrt{3}$. For $\beta = 2/\sqrt{3}$. The value of $f(\beta)$ is determined by both $2/\beta = 2/(2/\sqrt{3})=\sqrt{3}$ and $3- 2\alpha/\beta = 3 - \sqrt{3}(\sqrt{3}-1) =\sqrt{3}$. The value of $f(\cdot)$ in the interval $(2/\sqrt{3},\infty)$ is upper bounded by $\sqrt{3}$. Thus, $R\le \sqrt{3}$.
\fi
\end{proof}

\section{Restricted Assignment}
\label{sec:restricted_assign}

Recall that {\sc restricted assignment} is the special case of {\MM} in which $\pij \in \{ p_j, \infty\}$. We show that in this special case, the bound in Theorem \ref{thm:loaded_machines_oracle1} can be improved. 

\begin{theorem}
	\label{thm:loaded_machines_restricted_assign}
	$\mathsf{Oracle_{LM}}$ yields a $1.686$-approximation to the minimum makespan of restricted assignment. %, for any $m  > 1$.
\end{theorem}

\begin{proof}
Observe that since $\textbf{LPLST}(T)$ has a feasible solution then $p_j\le T$, $\forall j=1, \ldots , n$. We focus on Case 2.2 in the proof of Theorem~\ref{thm:loaded_machines_oracle1} ($\wtlp(T')$ has a feasible solution and $T' < 2 \alpha T$).
%We first note the slight difference in the notion of {\em long} jobs. Indeed, as the processing time of any job $j$ in a feasible schedule is $p_j \leq T$, we define the long jobs on machine $i$ as the set $J_i[\alpha T, T]$. Then, 
As before, the makespen of any machine $i \in \bml$ is bounded by $T' + \alpha T$. For machine $ i \in M_L$, we have that the total processing time on machine $i$ of all jobs assigned to this machine with processing time at most $T'-\alpha T$ is bounded by $2 (T'-\alpha T)$. Now, consider the long job assigned to machine $i \in M_L$. As the processing time of this job is at most $T$, we have that the completion time of this machine is at most $2(T' - \alpha T) + T$. Thus, the approximation ratio is bounded by 
\[
R= \max \{\frac{1}{\alpha}, 
    \min \{\frac{2}{\beta},
        \max \{1+\frac\alpha\beta,2 + \frac{1}{\beta}- \frac{2\alpha}{\beta}\}\}\},
\] 
for any $\beta\ge 1$, while we can optimize $\alpha\in (0,1)$. Fix $\alpha = \frac{\sqrt{33}-3}{4}$.
%We have that $\frac{1}{\alpha} < 1.46$. 
The maximum of the function $f(\beta) = \min \{\frac{2}{\beta},
\max \{1+\frac {\alpha}{\beta},2 + \frac{1}{\beta}- \frac{2\alpha}{\beta}\}\}$ for this fixed value of $\alpha$ is $1+\alpha = \frac{\sqrt{33}+1}{4}\approx 1.686$. %Then for $\beta \in [1, 2 \alpha)= [1,1.38)$, we have $f(\beta) \leq f(1)=1.69$, implying the same bound on $R$.

\end{proof}

\section{Graph Balancing}
\label{sec:graph_balance}

We now consider the {\GB} problem. Recall that this problem is a special case of {\sc restricted assignment} in which every job can be assigned to at most two machines.
%Thus, given an optimal solution, $\OPT$, for a graph balancing instance, we denote its value (i.e., the makespan of the corresponding schedule) by $\cmaxs$.
\begin{theorem}
\label{thm:graph_balancing_LR}
$\mathsf{Oracle_{LR}}$ yields a $\frac 12 + \frac 34\sqrt{2} \approx 1.561$-approximation to the graph balancing problem.
\end{theorem}
\begin{proof}
%Recall that~\cite{EKS2014} showed a 1.75-approximation for {\GB}.
The proof follows the proof of Theorem~\ref{thm:strong_oracle} while leveraging on the 1.75-approximation ratio for {\GB} given~\cite{EKS2014}. Substituting this improved approximation ratio instead of the 2-approximation for the general case, 
we get that $\mathsf{Oracle_{LR}}$ yields an approximation ratio $\max\{1+\alpha,\frac {1.75}{2\alpha}\}$.  The optimal $\alpha$ is given when $1+\alpha= \frac {1.75}{2\alpha}$, and $1+\alpha = \frac 12 + \frac 34  \sqrt{2} \approx 1.561$
\end{proof}

\begin{theorem}
\label{thm:graph_balancing}
$\mathsf{Oracle_{LM}}$ yields a $\frac58 \sqrt{7}\approx 1.6536$-approximation to the graph balancing problem.
\end{theorem}

\begin{proof}
%Recall that~\cite{EKS2014} showed a 1.75-approximation for the graph balancing problem. 
Define a parametrized version of \textbf{LP3} formulated in~\cite{EKS2014}, denoted $\textbf{LPEKS}(t)$, given in Figure~\ref{fig:lpeks}. Following~\cite{EKS2014}, let $E^B\subseteq E$ to be the set of edges with weight strictly larger than $t/2$. An edge in $E^B$ is called \emph{a big edge}. For an edge $e \in E$ with $v \in e$, let $x_{ev} \in \{ 0,1\}$ be an indicator for orienting $e$ towards $v$; that is, $\xev=1$ if $e$ is oriented towards $v$, and $\xev=0$ otherwise. In the linear programming relaxation, $\xev \geq 0$.

\begin{figure}[!ht]
\centering
\fbox{%
\begin{minipage}{\dimexpr 0.75\linewidth - 2\fboxsep - 2\fboxrule \relax}
{\small 
\begin{align}
\textbf{LPEKS}(t) 
\quad & x_{ev} +x_{eu} = 1  &&\quad \forall e=(u,v)\in E \label{rev:constraint_edges} \\
\quad & \displaystyle\sum_{e: v \in e} x_{ev}p_e  \leq t &&\quad \forall v\in V \label{rev:constraint_vertices} \\
\quad & \displaystyle\sum_{e \in E^{B}: v \in e} x_{ev} \leq 1 &&\quad \forall v\in V \label{rev:constraint_big_edges} \\
\quad & x_{ev} \geq 0  &&\quad \forall  e,v: v \in e \nonumber 
\end{align}
}
\end{minipage}
}
\caption{The LP formulation $\textbf{LPEKS}(t)$}
\label{fig:lpeks}
\end{figure}

Find the minimum $t$ for which $\textbf{LPEKS}(t)$ is feasible. It follows from~\cite{EKS2014} that we can compute a solution with maximum load $1.75t$. Note that since $t$ is just a lower bound on $\cmaxs$,
%the value of the optimal solution, 
it may well be that the optimal solution $\OPT$ has two edges in $E^B$ incoming to the same vertex (which cannot happen if $t$ upper bounds $\cmaxs$). 
%the value of the optimal solution).

Fix $r>1$ (to be computed later), assume that $rt$ is an upper bound on $\cmaxs$. Note that if this assumption is wrong the $1.75t$ solution computed using the algorithm in~\cite{EKS2014} yields a $1.75/r$ approximation. From now on (until we fix the value of $r$) we assume that $rt$ is indeed an upper bound on the value of the optimal solution.

We ask the oracle to specify the set of vertices (machines) that have an incoming edge (job) of weight at least $hrt$, for some $h>\frac 12$ to be specified later. We call all the edges of weight at least $hrt$ \emph{huge} edges; denote this set of edges by $E^H$. We call the edges of weight larger than $\frac 12 rt$ \emph{large} edges, and denote the set of these edges by $\eob$. Note that $E^H\subseteq \eob$.

Let $V^H$ be the set output by the oracle. If a huge edge $e\in E^H$ has only one endpoint $v\in V^H$ it must be oriented towards $v$. We fix this orientation, remove $e$ from the instance, and add $p_e$ to the weight $q_v$ of $v$ (initially, $q_v=0$). We iterate until all the huge edges with only one endpoint in $V^H$ are removed.

Now, we continue to orient edges as follows. As long as there is a large edge $e=(u,v)\in \eob$ such that one of its endpoints, say $u$, already has a fixed incoming large edge, we orient $e$ towards $v$, remove $e$ from the instance, and add $p_e$ to $q_v$. If $v$ already has a fixed incoming large edge as well, or if $v\in V^H$ but $e\notin E^H$, or if $e\in E^H$ but $v\notin V^H$, then $rt$ is not an upper bound on $\cmaxs$.

Let $\hat{E}$ be the set of remaining edges, and let $\hat{E}^L$ and $\hat{E}^H$ be the remaining large and huge edges. Also, let $\hat{V}^H\subseteq V^H$ be the vertices with no fixed incoming edges. Note that every edge $e\in\hat{E}^H$ has both its endpoints in $\hat{V}^H$ and  $|\hat{E}^H|=|\hat{V}^H|$. 
It follows that each connected component of $\hat{G}^H(\hat{V}^H,\hat{E}^H)$ is a pseudo tree. Moreover, since $\hat{G}^H$ has no degree 1 vertices, each connected component is a cycle, and $\hat{G}^H$ is a collection of disjoint cycles. 
Note that no vertex in $V\setminus \hat{V}^H$ can have an incoming edge $e\in \hat{E}$ with $p_e>rt-q_v$. Also, no vertex in $\hat{V}^H$ can have an incoming edge $e\in\hat{E}\setminus \hat{E}^H$ with $p_e>rt-hrt$.

We define the following linear program $\widehat{\textbf{LPEKS}}$, given in \ref{fig:hatlpeks}. In this LP, in addition to a variable for every edge endpoint,
there is a variable $y_c$ for every cycle $c$ in $\hat{G}^H$ that identifies a global direction of the cycle, say $y_c =1$ is clockwise direction and $y_c =0$ is counterclockwise direction. For an edge $e=(u,v)\in \hat{E}^H$, we denote the cycle to which it belongs by $c(e)$, and assume that $u$ (first vertex in the pair) precedes $v$ (second vertex in the pair) in the clockwise direction of $c_e$.

\begin{figure}[!ht]
\centering
\fbox{%
\begin{minipage}{\dimexpr 0.98\linewidth - 2\fboxsep - 2\fboxrule \relax}
{\small 
\begin{align}
\widehat{\textbf{LPEKS}} 
\quad & x_{ev} +x_{eu} = 1  &&\quad \forall e=(u,v)\in \hat{E}\setminus \hat{E}^H \label{rev:constraint_hat_edges} \\
\quad & \displaystyle q_v+\sum_{e \in \hat{E}: v \in e} x_{ev}p_e  \leq rt &&\quad \forall v\in V\setminus\hat{V}^H \label{rev:constraint_vertices_nocycle} \\
\quad & \displaystyle x_{ev}=0 &&\quad \forall v\in V\setminus\hat{V}^H \, \forall e \in \hat{E} : v \in e \quad \text{s.t.} \quad p_e>rt-q_v \label{rev:constraint_fixed_edges_1} \\
\quad & \displaystyle \sum_{e \in \hat{E}\setminus \hat{E}^H : v \in e} x_{ev}p_e + 
\sum_{e=(u,v)\in \hat{E}^H} y_{c(e)}p_e &&+ \sum_{e=(v,u)\in \hat{E}^H} (1-y_{c(e)}) p_e 
\leq rt \quad  \forall v\in \hat{V}^H \label{rev:constraint_vertices_cycle} \\
\quad & \displaystyle x_{ev}=0 &&\quad \forall v\in \hat{V}^H  \, \forall e \in \hat{E}\setminus \hat{E}^H : v \in e \quad \text{s.t.}\quad p_e>rt(1-h)  \label{rev:constraint_fixed_edges_2} \\
\quad & \displaystyle\sum_{e\in \hat{E}^{L}: v \in e} x_{ev}  \leq 1 &&\quad \forall v\in V \label{rev:constraint_big_edges_hat} \\
\quad & x_{ev} \geq 0  &&\quad \forall e\in \hat{E}: v \in e \nonumber \\
\quad &  0 \leq y_{c} \leq 1  &&\quad \text{for all cycles } c \text{ in }\hat{G}^H \nonumber 
\end{align}
}
\end{minipage}
}
\caption{The LP formulation $\widehat{\textbf{LPEKS}}$}
\label{fig:hatlpeks}
\end{figure}

Constraints~(\ref{rev:constraint_hat_edges}) ensure that all remaining edges are oriented. Constraints~(\ref{rev:constraint_vertices_nocycle}) guarantee that the total load on each vertex $v \notin \hat{V}^H$ is at most $rt$, the upper bound we impose on $\cmaxs$. Constraints~(\ref{rev:constraint_fixed_edges_1}) forbid orienting an edge $e$ towards an endpoint $v$ if $p_e > rt-q_v$. Constraints~(\ref{rev:constraint_vertices_cycle}) govern the orientation of cycles whose vertices all belong to $\hat{V}^H$. They ensure that each such cycle is assigned a single orientation, either clockwise or counterclockwise, and that the total load on each of its vertices does not exceed $rt$.
Constraints~(\ref{rev:constraint_fixed_edges_2}) concern the vertices $v \in \hat{V}^H$. Each such vertex $v$ already has load at least $rth$ due to an incoming huge edge on a cycle containing $v$. Therefore, the additional load that can be assigned to $v$ is at most $rt(1-h)$. Finally, constraints~(\ref{rev:constraint_big_edges_hat}) ensure that at most one large edge is oriented towards each vertex.

Next, we round the fractional solution of $\widehat{\textbf{LPEKS}}$ as follows. First, we round every variable $y_c$; that is, if $y_c\ge 1/2$ we orient the cycle clockwise; otherwise, we orient it counterclockwise. This takes care of all the edges in $\hat{E}^H$. The rounding error due to the edges in $\hat{E}^H$ is at most $\frac 12 (rt-hrt)$. For the rest of the edges we use a modification of the rounding technique of~\cite{EKS2014}. Note that while the upper bound on the optimal solution is $rt$ we do not have any edge of weight greater than $hrt$. As shown in Lemma~\ref{lem:EKS2014} below, in this case a modification of the algorithm of~\cite{EKS2014} guarantees an approximation ratio of $\frac 54 + \frac h2$. Since at most one edge incoming to a vertex is rounded up by the algorithm of~\cite{EKS2014}, the additional rounding error of vertices in $\hat{V}^H$ due to incoming edges not in $\hat{E}^H$ is bounded by $(rt-hrt)$.

In summary, assuming that $rt$ and $t$ are upper and lower bounds on $\cmaxs$, we get an approximation ratio of $\max\{\frac 52 r - \frac 32 hr ,\frac 54 r+ \frac 12 {rh}\}$.  Indeed, the left term bounds the total load on vertices in $\hat{V}^H$ compared to the lower bound of $t$, i.e., $\left[rt + \frac{1}{2} (rt -hrt) + rt -hrt\right])\cdot \frac 1t= \frac{5}{2}r - \frac{3}{2} hr$. The right term follows from the bound in Lemma~\ref{lem:EKS2014}, applied to vertices $v \notin \hat{V}^H$. Since the assumption that $rt$ is an upper bound on the optimal solution may be wrong, we also need to include the term $\frac 7{4r}$, which is the approximation ratio in case the assumption is wrong. To optimize the approximation ratio, we need to find the minimum over $r\ge 1$ and $\frac 12 \le h \le 1$ of  $\max\{\frac 7{4r}, \frac 52 r - \frac 32 hr ,\frac 54 r+ \frac 12 {rh}\}$. We first equate the second and third terms to find the optimal $h=\frac 58$. Then we equate the first term with the second and third ones to get $r=\frac {2\sqrt{7}}5$, and approximation ratio of $\frac{5\sqrt{7}}8 \approx 1.6536$.
\end{proof}

We now show how to modify the algorithm of~\cite{EKS2014} to achieve an improved approximation factor in case the instance does not include very large weights relative to $\cmaxs$, as required in the proof of Theorem~\ref{thm:graph_balancing}.

\begin{lemma}
\label{lem:EKS2014}
For a given instance of the graph balancing problem $G(V,E,p)$ and a given $t>0$, suppose that $\emph{\textbf{LPEKS}}(t)$ is feasible. If $\forall e\in E,\ p_e < ht$, for some $h\in(\frac 12,1]$, then there exists a polynomial $\left( \frac 54+\frac h2 \right)$-approximation algorithm to this instance.
\end{lemma}
\begin{proof}
    We follow the algorithm given in~\cite{EKS2014} but instead of assigning a big edge to a leaf if this causes an increase of $\frac 34 t$ in the load we fix some $s\in (\frac 12 ,1]$ and set the increase threshold to $sht$. The optimal $s$ is given by equating the two possible increases in the load; that is, $sht = \frac 12 t + (1-s)ht$. It follows that $s=\frac 12+\frac 1{4h}$, and the increase in the load is $\frac h2 +\frac 14$.
\end{proof}

%%%%%%%%%%%%%%%%%%%%%%%%%%%%%%%%%%%%%%%%%%%%%%%%%%%%%%%%%%
\
%\input{GAP}

%%%%%%%%%%%%%%%%%%%%%%%%%%%%%%%%%%%%%%%%%%%%%%%%%%%%%%%%%%%%%
\section{The Santa Claus Problem}
\label{sec:santa_claus}
In this section we present our results for the {\SC} problem.
%of maximizing the minimum load of unrelated machines. 
%An instance consists of a set $M$ of $m$ machines and a set $J$ of $n$ jobs. Let $\pij \in {\mathbb{Z}}^+$ denote the processing time of job $j$ on machine $i$, for $1 \le j \le n$, $1 \leq i \leq m$. the goal is find an allocation of jobs to machines that maximizes the minimum load (total processing time) over the set of machines $M$. 

\iffalse
=======> I DON'T THINK THIS IS NEEDED
This max-min fair allocation problem can be relaxed to the following fractional LP. 
\begin{align}
\textbf{SCLP} 
    \quad \max \quad & \omega \nonumber \\
    \quad \text{s.t.} \quad & \sum_i x_{ij} = 1 \quad \forall j\in J \label{eq:assign} \\
    & \sum_j \pij x_{ij} \ge \omega \quad \forall i\in M \label{eq:utility} \\
    & 0 \le x_{ij} \le 1 \quad \forall i\in M,\, j\in J \nonumber
\end{align}
%Let $C_{min}^*$ denote the optimal value of this linear program. 
\fi
    
\subsection{Warm-up: Max-Min Fair Allocation with the Long Jobs Schedule Oracle}
Suppose we have an oracle that observes an optimal allocation of a given instance. Let $\FOPT$ be an optimal solution and let $C_{min}^*$ be the value of $\FOPT$. For a parameter $\alpha \in (0,1)$, let $J_L\subseteq J$ be the subset of jobs with processing time at least $\alpha \cdot C_{min}^*$ in the optimal assignment. The strong oracle reveals $J_L$ and for each $j \in J_L$ it also reveals the machine $i(j)$ to which $j$ is assigned in $\FOPT$. (Note that $C_{min}^*$ is not revealed by the oracle.)

\begin{theorem}
\label{thm:SCstrong}
$\mathsf{Oracle_{LS}}$ yields a $\frac 1{1 - \alpha}$-approximation to the Max-Min fair allocation problem.
\end{theorem}
\begin{proof}
    Let $M_L$ denote the set of machines to which the jobs in $J_L$ are assigned. Let $J(i)$ denote the set of long jobs assigned on machine $i \in M_L$. Consider the LP $\widehat{\textbf{SCLP}}$ defined as follows.

\begin{figure}[htbp]
\centering
\fbox{%
\begin{minipage}{\dimexpr 0.75\linewidth - 2\fboxsep - 2\fboxrule \relax}
{\small
\begin{align}
\widehat{\textbf{SCLP}}
\quad \max \quad & \omega \nonumber \\
\quad \text{s.t.} \quad & \sum_i x_{ij} \le 1 &&\quad \forall j\in J   \label{eq:assign_modified} \\
\quad & x_{i(j),j} = 1  &&\quad \forall j \in J_L \label{SCfixed}\\
\quad & x_{i,j} = 0  &&\quad \forall j \in J \setminus J_L\text{ s.t. } p_{ij} \ge \alpha \cdot \omega \label{SCnotassign}\\
& \sum_{j \in J \setminus J_L} \pij x_{ij} \ge \omega -\sum_{j\in J(i)} \pij &&\quad \forall i\in M_L\label{eq:utility_big} \\
& \sum_{j \in J \setminus J_L} \pij x_{ij} \ge \omega &&\quad \forall i\in M\setminus M_L\label{eq:utility_small} \\
& 0 \le x_{ij} &&\quad \forall i\in M,\, j\in J\setminus J_L \nonumber
\end{align}
}
\end{minipage}
}    
\caption{The LP formulation $\widehat{\textbf{SCLP}}$}
\label{fig:hatSC1}
\end{figure}

%In Constraint(~\ref{SCfixed}) we fix the assignment of the long jobs according to the oracle's advice.
Constraints~(\ref{eq:utility_big}) and~(\ref{eq:utility_small}) guarantee that  
every machine is assigned a total load of at least $\omega$. For machines in $M_L$, jobs with processing time less than $\alpha \cdot \omega$ should complete to $\omega$ the load generated by the jobs in $J(i)$, while for machines not in $M_L$, the whole load is due to jobs not in $J_L$. Constraints~(\ref{SCnotassign}) guarantee that jobs not in $J_L$ are assigned to machines on which their processing is less than $\alpha \cdot \omega$.  

Let $\omega^*$ be the value of an optimal solution to $\widehat{\textbf{SCLP}}$. It is not difficult to see that $\FOPT$ induces a feasible solution of $\widehat{\textbf{SCLP}}$. Therefore, $\omega^* \ge C^*_{min}$. Also, the rounding described in Theorem 7 in~\cite{BD05} can be applied to $\widehat{\textbf{SCLP}}$. It follows that any basic feasible solution to this LP can be rounded to an integer solution such that the load of machine decreases by at most the value of the largest fractionally assigned job.

Since all the non-integral variables refer to jobs and machines for which $p_{ij} < \alpha \cdot \omega$, the rounding procedure guarantees a minimum utility of at least:
    $$ \omega^* - \max_{ij: 0<x_{ij}< 1} \pij > \omega^* - \alpha \cdot \omega^*\ge C_{min}^* - \alpha \cdot C_{min}^* = (1 - \alpha)C_{min}^*. $$
\end{proof}

\subsection{Max-Min Fair Allocation with the Long Job Machines Oracle}
Suppose we have an oracle that observes an optimal allocation of a given instance. Let $\FOPT$ be an optimal solution and let $C_{min}^*$ be the value of $\FOPT$.
For a parameter $\alpha \in (0,1)$, let $M_L\subseteq M$ be the subset of machines that execute jobs with processing time at least $\alpha C_{min}^*$ in 
$\FOPT$. $\mathsf{Oracle_{LM}}$ reveals the set $M_L$. We present an algorithm that uses $\mathsf{Oracle_{LM}}$ to get a $\max(\frac 1\alpha, \frac 1{1- \alpha})$-approximation. Note that for $\alpha=\frac 12$, this is a $2$-approximation.  
\begin{theorem}
\label{thm:SCweak}
$\mathsf{Oracle_{LM}}$ yields a $\max(\frac 1\alpha, \frac 1{1- \alpha})$-approximation to the Max-Min fair allocation problem.
\end{theorem}
\begin{proof}
Consider the linear program $\widetilde{\textbf{SCLP}}$.

\begin{figure}[htbp]
\centering
\fbox{%
\begin{minipage}{\dimexpr 0.75\linewidth - 2\fboxsep - 2\fboxrule \relax}
{\small
\begin{align}
\widetilde{\textbf{SCLP}}
\quad \max \quad & \omega \nonumber \\
\quad \text{s.t.} \quad & \sum_i x_{ij} = 1 &&\quad \forall j\in J   \label{eq:assign_modified_loaded} \\
\quad & x_{ij} = 0 &&\quad \forall i\in M \setminus M_L,\, p_{ij}\ge \alpha \omega \label{eq:nolong}\\
\quad & x_{ij} = 0 &&\quad \forall i\in M_L,\, p_{ij} < \alpha \omega \label{eq:noshort}\\
& \sum_j \pij x_{ij} \ge \omega &&\quad \forall i\in M\setminus M_L \label{eq:utility_small_loaded} \\
& \sum_{j:p_{ij} \ge \alpha \omega} x_{ij} =1 &&\quad \forall i\in M_L \label{eq:utility_big_loaded} \\
& 0 \le x_{ij} &&\quad \forall i\in M,\, j\in J \nonumber
\end{align}
}
\end{minipage}
}    
\caption{The LP formulation $\widetilde{\textbf{SCLP}}$}
\label{fig:hatSC2}
\end{figure}

Constraint~(\ref{eq:nolong}) guarantees that the processing time of a job assigned to a machine not in $M_L$ is less than $\alpha\omega$. 
Constraint~(\ref{eq:utility_small_loaded}) guaranties that  
every machine not in $M_L$ is assigned a total load of at least $\omega$. By Constraint~(\ref{eq:nolong}) this load is generated by jobs with processing time less than $\alpha\omega$. Constraint~(\ref{eq:utility_big_loaded}) guaranties that  
every machine in $M_L$ is assigned exactly one job with processing time at least $\alpha\omega$. By Constraint~(\ref{eq:noshort}) this is the only job on the machine. 

Let $\omega^*$ be the value of an optimal solution to $\widetilde{\textbf{SCLP}}$. Note that the optimal allocation $\FOPT$ implies a feasible solution of $\widetilde{\textbf{SCLP}}$. This is because each machine $i \notin M_L$ is allocated only jobs with processing time less than $\alpha C_{min}^*$ that must sum to at least $C_{min}^*$ (constraints~(\ref{eq:utility_small_loaded})), and each machine $i \in M_L$ is allocated at least one job with processing time $\ge \alpha C_{min}^*$ (constraints~(\ref{eq:utility_big_loaded})). Therefore, $\omega^* \ge C_{min}^*$.

We round the fractional solution of $\widetilde{\textbf{SCLP}}$ using the techniques of~\cite{LST90}. First, we fix each variable $x_{ij}$ that is integral in the solution of $\widetilde{\textbf{SCLP}}$. Next, we consider the bipartite graph $\tilde{G}(\tilde{J},\tilde{M},\tilde{E})$, where $\tilde{J}$ is the set of all jobs $j$ such that there exists $i\in M$ for which $x_{ij}$ is fractional in the solution of $\widetilde{\textbf{SCLP}}$, $\tilde{M}$ is the set of all machines $i$ such that there exists $j\in J$ for which $x_{ij}$ is fractional in the solution of $\widetilde{\textbf{SCLP}}$, and $(i,j)\in \tilde{E}$ iff $x_{ij}$ is fractional. As in~\cite{LST90}, $\tilde{G}$ is a pseudo-forest. Also, Lemma 8 in~\cite{BD05} can be applied to find a subset $S\subseteq\tilde{E}$ such that each $j\in \tilde{J}$ is the endpoint of exactly one edge in $S$ and each $i\in \tilde{M}$ is the endpoint of at least $deg(i)-1$ edges in $S$ (where $deg(i)$ is the degree of $i$ in $\tilde{G}$). 
The set $S$ together with the variables that were fixed to $1$ induces a job allocation. We analyze the load of each machine. Recall that $\omega^* \ge C_{min}^*$: 
%in this job allocation is at least $\frac 12 C_{min}^*$. To see this consider two cases:
\par\noindent\textbf{Case 1:} The machine $i\in M_L$. If there exists $j\in J$ such that $x_{ij}=1$ then it must be that $\pij\ge \alpha\omega^*\ge \alpha C_{min}^*$. Otherwise, $i\in \tilde{M}$, and $deg(i)\ge 2$ since Constraint~(\ref{eq:utility_big_loaded}) holds. It follows that at least one edge, touching $i$, say $(i,j')$, is in $S$. Since $\pij'\ge \alpha C_{min}^*$ the load is also at least $\alpha C_{min}^*$.
\par\noindent\textbf{Case 2:} The machine $i\in M\setminus M_L$. Let $J(i)\subseteq J$ denote the set of jobs that are assigned to $i$, either fractionally or integrally, in the solution. Note that for each $j\in J(i)$, $\pij < \alpha \omega^*$. Also, $\sum_{j\in J(i)} \pij \ge \omega^*$ due to Constraint~(\ref{eq:utility_small_loaded}). Since all the jobs $j\in J(i)$, for which $x_{ij}=1$ are assigned to $i$, and all the jobs $j\in J(i)$, for which $x_{ij}<1$, except for at most one job, are also assigned to $i$, the load of machine $i$ is at least $(1-\alpha)\omega^*\ge (1-\alpha)C_{min}^*$. 

We conclude that the load on every machine is at least $\min(\alpha, 1- \alpha)C_{min}^*$.
\end{proof}

\section{Hardness of Approximation with Oracle Access}
\label{sec:hardness_oracles}
While the oracles introduced in this paper are highly beneficial, enabling us to bypass classical bottlenecks and improve upon the best-known approximation ratios, their power is inherently limited. In this section, we explore these theoretical boundaries. We prove that the underlying scheduling and allocation problems remain not only NP-hard, but strictly hard to approximate, even when the algorithm is augmented with the information provided by these oracles.

We first mention a simple straightforward result, regarding oracles we have not considered in this paper. The proof is based on the known hardness of approximation of restricted scheduling, and  an addition of tiny dummy jobs. 

\begin{theorem}
It is NP-hard to approximate the makespan of $R||C_{max}$ within a factor less than $\frac 32$, even with an oracle that reveals $\cmaxs$, the actual jobs' lengths, as well as the number of jobs assigned to every machine in some optimal solution.
\end{theorem}

Next, we show that \textsc{Graph Balancing} remains NP-hard and hard to approximate, even with an access to $\mathsf{Oracle_{LM}}$, as long as the oracle reveals information about edges with weight more than $\alpha \cmaxs$, for any $\alpha \le \frac 12$. Since \textsc{Graph Balancing} is a restricted class of $R||C_{max}$, our result clearly captures also the {\MM} problem.

\begin{theorem}
\label{thm:hardMM}
For any $\alpha \le \frac 12$, it is NP-hard to approximate the minimum makespan in a \textsc{Graph Balancing} problem within factor less than $\frac {3} 2$, even with an oracle that reveals the value of $\cmaxs$ and the set of vertices that have an incoming edge of weight more than $\alpha\cmaxs$ in some optimal solution.
\end{theorem}

\begin{proof}
Assume first that $\alpha=\frac 12$. Our reduction is from $3$-SAT. Given a $3$-SAT formula, we construct an instance of \textsc{Graph Balancing} such that a satisfying assignment exists if and only if there is an orientation with a maximum load of $\cmaxs = 2$. Moreover, the set of vertices that receive an edge of weight more than $1$ is independent of the solution.

Let $0< \epsilon < \frac 1 3$.
For each clause $C = (\ell_1 \lor \ell_2 \lor \ell_3)$, we create a single \textbf{clause vertex} $v_C$. It has $3$ incident edges connecting to the respective variable gadgets, each representing a literal $\ell_i$. All edges connecting a clause vertex and a variable gadget have  weight $1-\epsilon$.

For each variable $x$, let $k = \max(\text{count}(x), \text{count}(\bar{x}))$ denote the maximum number of times either the literal $x$ or its negation $\bar{x}$ appears in the formula. The {\bf variable gadget} of $x$ consists of a cycle of length $2k$. See Figure~\ref{fig:variablegadget} for an example with $k=3$. 
\begin{itemize}
    \item The vertices along the cycle alternate based on the literal they host: $v^1_x, v^1_{\bar{x}}$, $v^2_x, v^2_{\bar{x}}, \dots, v^{k}_x, v^k_{\bar{x}}$. 
    \item The cycle edges alternate in weight between $2$ and $1+\epsilon$. They are aligned such that moving in clockwise direction, an edge of weight $1+\epsilon$ always enters an $x$-node, and an edge of weight $2$ always enters a $\bar{x}$-node.
    \item Each literal occurrence in the formula contributes an edge of weight $1-\epsilon$ connecting an available vertex of its type in the cycle and the corresponding clause vertex.
    \item Note that if the literal counts are unequal (e.g., $\text{count}(x) = 2$ and $\text{count}(\bar{x}) = 3$, meaning $k=3$), we still build a full $2k$-node cycle with alternating node types. $|\text{count}(x) -\text{count}(\bar{x})|$ vertices corresponding to the literal with fewer occurrences will not be connected to a clause vertex.
\end{itemize}

\begin{figure}[!ht]
  \centering
  \includegraphics[width=0.4\textwidth]{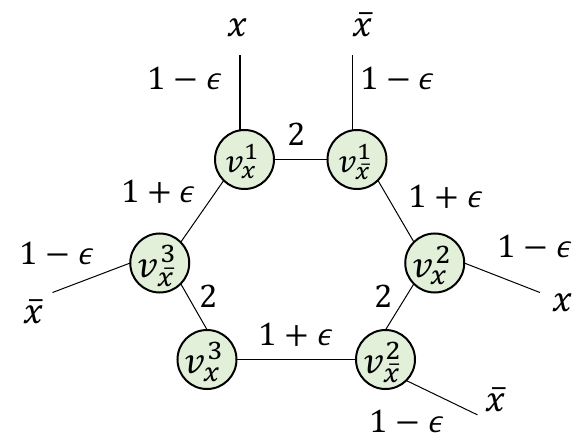}
  \caption{The variable gadget for variable $x$ in a $3$-SAT formula in which $x$ appears $2$ times and $\bar x$ appears $3$ times. The five outputs are connected to the corresponding clause vertices.}
  \label{fig:variablegadget}
\end{figure}

\begin{claim}
\label{cl:reductionGB}
A $3$-SAT formula is satisfiable if and only if the corresponding graph balancing instance has an orientation with a maximum load of $2$.
\end{claim}

%Moved to App
\begin{proof}
Assume a satisfying truth assignment exists. We orient the graph as follows:
\begin{enumerate}
    \item For each clause, at least one literal evaluates to \texttt{True}. We orient the external edge corresponding to this true literal out of the clause vertex $v_C$ (into its variable gadget). The remaining edges are oriented into $v_C$. Since at most two edges enter $v_C$, its total incoming load is at most $2 \cdot (1-\epsilon) < 2$.
    \item If a variable $x$ is \texttt{True}, we orient its cycle clockwise. This means internal edges of weight $2$ are oriented from $v^i_x$ to $v^i_{\bar{x}}$, and internal edges of weight $1+\epsilon$ are oriented from $v^i_{\bar{x}}$ to $v^{i+1}_{x}$. Furthermore, all external edges representing the negated literal $\bar{x}$ are directed out of the cycle (into their respective clauses), while the external edges for $x$ are directed into the cycle.

    We evaluate the incoming load for any cycle vertex $u$:
    $(i)$ If $u$ is incident to an external edge directed out of the cycle, it is entered by an edge of weight $2$. 
    $(ii)$ If $u$ is incident to an external edge (of weight $1 - \epsilon$) directed  into the cycle, it is entered also by a cycle edge of weight $1+\epsilon$.  The total load into $u$ is $(1+\epsilon) + (1-\epsilon) = 2$.         
    If a variable is \texttt{False}, the analysis is symmetric (orienting the cycle counter-clockwise). Thus, the maximum load across all the cycle vertices is exactly $2$.
 \end{enumerate}

For the other direction of the proof, assume there exists an orientation with a maximum load of $2$. 

Each vertex $u$ in a variable gadget is incident to a cycle edge of weight $2$ and a cycle edge of weight $1+\epsilon$. The cycle must be oriented consistently either entirely clockwise or entirely counter-clockwise, as otherwise some vertex would receive a load of at least $2 + (1+\epsilon) > 2$. 

For any clause vertex $v_C$, if all three incident edges of weight $1-\epsilon$ were oriented into it, its total load would be $3 - 3\epsilon > 2$. Given that the maximum load is $2$, every clause vertex must have at least one incident edge oriented out of it (into a variable gadget).

The orientation induces a truth assignment: if a variable's cycle is oriented such that vertices $v_i^x$ are entered by edges of weight $(1+\epsilon)$, then its positive literal edges can also enter the cycle, and we set it to \texttt{True}; otherwise, we set it to \texttt{False}. Since the cycles are oriented entirely clockwise or counter-clockwise there are no conflicts in the literal assignments, ensuring the resulting truth assignment is perfectly consistent. Since every clause has at least one edge out of it, every clause is satisfied.
\end{proof}

Next, we show that an oracle that reveals which vertices receive a job of weight strictly greater than $\frac 12 \cmaxs$ does not enable an approximation better than $\frac 3 2 -\epsilon$. 
First note that all edges adjacent to a clause vertex have a weight of $1-\epsilon < 1$. Since $\cmaxs \ge 2$, the oracle will not inform any clause vertex. 

If $\cmaxs=2$, then every variable cycle must be oriented in one direction. Since all cycle edges have weight $>1$, every cycle vertex has exactly one incoming edge of weight $>1$ regardless of the optimal assignment (and the corresponding truth assignment). The oracle will therefore inform  all variable vertices.
Since this information is independent of the solution, and $3$-SAT is NP-hard even if it is known that a truth assignment exists, the oracle does not provide any useful information.

Finally, if $\cmaxs>2$, then $\cmaxs \ge 3-\epsilon$ implying a gap of $\frac {3- \epsilon}{2}$. Since $\epsilon$ can be arbitrarily small, no algorithm can guarantee approximation ratio lower than $\frac 32$.

To extend the proof also for $\alpha < \frac 12$, set $\epsilon < \min(\frac 1 3, 1 -2\alpha)$ and consider the same reduction but with a 3SAT formula that has the property that it has no satisfying assignment in which there is a clause that all its literals are satisfied. It is well know that the problem remains NP-Hard also when the input is restricted to such instances~\cite{Sch78}. Note that for such instances at least one edge must be incoming to the clause vertex and thus all vertices have an incoming edge of weight at least $1 -\epsilon > \alpha \cmaxs$ in some optimal solution. Thus, this additional information does not enable an approximation better than $\frac 3 2 -\epsilon$. 
\end{proof}

We proceed to show that \textsc{Graph Balancing} remains NP-hard and hard to approximate, even with an access to $\mathsf{Oracle_{LM}}$, as long as the oracle reveals information about edges with weight more than $\alpha \cmaxs$, for any $\frac 12<\alpha <1$. 

\begin{theorem}
\label{thm:hardMMalpha}
It is NP-hard to approximate the minimum makespan in a \textsc{Graph Balancing} problem within factor less than $2-\alpha$, even with an oracle that reveals the value of $\cmaxs$ and the set of vertices that have an incoming edge of weight more than $\alpha \cmaxs$ in some optimal solution, for any $\frac 12<\alpha <1$.
\end{theorem}

\begin{proof}
We use a similar construction as in the proof of Theorem~\ref{thm:hardMM}. Let $\alpha' = \alpha + \epsilon$, for some $0 < \epsilon < \frac{1 - \alpha}{2}$. The differences are as follows: (i) The cycle edges in the variable gadgets alternate in weight between $2$ and $2\alpha'$. (ii) All edges connecting a clause vertex and a variable gadget have  weight $2-2\alpha'<2\alpha$. (iii) A self-loop of weight  $4\alpha-2<2\alpha$ is attached to every clause vertex. Following the proof of Theorem~\ref{thm:hardMM} the graph balancing instance has an orientation with a maximum load of $2$ if and only if the 3SAT instance has a satisfying assignment. Note that every cycle edge in the variable gadgets has weight either $2$ or $2\alpha' = 2\alpha + 2\epsilon > 2\alpha$, and for every clause vertex, all its incident edges are of weight $2 - 2\alpha' < 2 - 2\alpha < 2\alpha$ (since $\alpha > \frac{1}{2}$), or $4\alpha' - 2 = 4\alpha - 2 + 4\epsilon < 4\alpha - 2 + 2(1 - \alpha) = 2\alpha$. Thus, in any solution with $\cmaxs=2$ every vertex in the variable gadgets must have an incoming edge of weight at least $2\alpha$. Clearly, none of the clause vertex can have an incoming edge of weight at least $2\alpha$.  It follows that the oracle informs all the vertices in variable gadgets, and none of the clause variables, deeming the oracle useless. 

If $\cmaxs>2$, then $\cmaxs \ge 4-2\alpha$ implying a gap of $2-\alpha$.

\end{proof}

%%%%%%%%%%%%%%%%%%%%%%%%%%%%%%%%%%%%%%%%%%%%%%%%%%%%%%
Finally, we show a hardness result for approximating the {\SC} problem (max min fair allocation) even if the algorithm can consult $\mathsf{Oracle_{LM}}$. Specifically, we show that no algorithm can achieve better that $2$-approximation even if it can consult $\mathsf{Oracle_{LM}}$. Since our algorithm that consults $\mathsf{Oracle_{LM}}$ achieves 2 approximation (when $\alpha$ is fixed to 2), this lower bound is tight. Our hardness refers to a restricted class of {\SC} corresponding to {\GB}, that is, every job has a fixed length and can only be assigned to one of at most two machines. In graph balancing terms, our goal is therefore to direct the graph edges such that the minimal weight entering a vertex is maximized.  

\begin{theorem}
\label{thm:hardSC}
For any $\alpha \le \frac 12$, it is NP-hard to approximate the {\MMGB} problem within a factor less than $2$, even with an oracle that reveals the value of $\cmins$ and the set of vertices that have an incoming edge of weight more than $\alpha\cmins$ in some optimal solution. 
\end{theorem}

\begin{proof}
Our reduction is from {$3$-SAT}. Given a $3$-SAT formula, we construct an instance of {\MMGB} such that a satisfying assignment exists if and only if there is an orientation with a minimum load of $C_{min}^* = 2$. Moreover, the set of vertices that receive an edge of weight more than $1$ is independent of the solution.

Let $0< \epsilon < \frac 1 3$.
For each clause $C = (\ell_1 \lor \ell_2 \lor \ell_3)$, we create a single \textbf{clause vertex} $v_C$. It has a self-loop of weight $1$, and three incident edges connecting it to the variable gadgets corresponding to literals $\ell_1$, $\ell_2$, and $\ell_3$. All edges connecting a clause vertex and a variable gadget have  weight $1+\epsilon$.

For each variable $x$, let $k = \max(\text{count}(x), \text{count}(\bar{x}))$ denote the maximum number of times either the literal $x$ or its negation $\bar{x}$ appears in the formula. The {\bf variable gadget} of $x$ consists of a cycle of length $2k$. See Figure~\ref{fig:variablegadgetSC} for an example with $k=3$. 

\begin{itemize}
    \item The vertices along the cycle alternate based on the literal they host: $v^1_x, v^1_{\bar{x}}$, $v^2_x, v^2_{\bar{x}}, \dots, v^{k}_x, v^k_{\bar{x}}$. 
    \item The cycle edges alternate in weight between $2$ and $1-\epsilon$. They are aligned such that moving in clockwise direction, an edge of weight $1-\epsilon$ always enters an $x$-node, and an edge of weight $2$ always enters a $\bar{x}$-node.
    \item Each literal occurrence in the formula contributes an edge of weight $1+\epsilon$ connecting an available vertex of its type in the cycle and the corresponding clause vertex.
    \item If the literal counts are unequal (e.g., $\text{count}(x) = 2$ and $\text{count}(\bar{x}) = 3$, meaning $k=3$), we still build a full $2k$-node cycle with alternating node types. $|\text{count}(x) -\text{count}(\bar{x})|$ vertices corresponding to the literal with fewer occurrences have a self loop of weight $1 +\epsilon$.
\end{itemize}

\begin{figure}[!ht]
  \centering
  \includegraphics[width=0.4\textwidth]{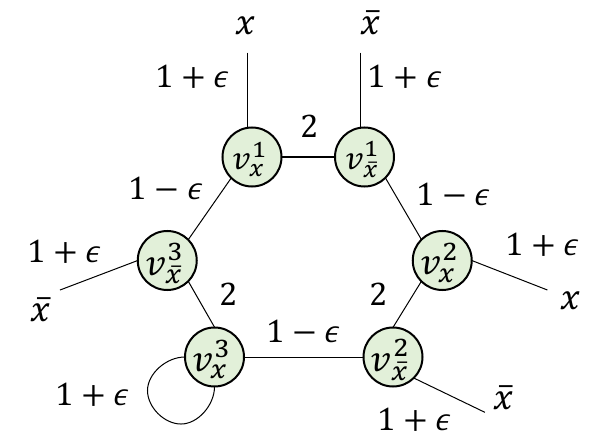}
  \caption{The variable gadget for variable $x$ in a $3$-SAT formula in which $x$ appears $2$ times and $\bar x$ appears $3$ times. The five outputs are connected to the corresponding clause vertices. One output is a self-loop.}
  \label{fig:variablegadgetSC}
\end{figure}

\begin{claim}
An {\sc $3$-SAT} formula is satisfiable if and only if the corresponding Santa Claus instance has an orientation with a minimum load of $2$.
\end{claim}

\begin{proof}
Assume a satisfying truth assignment exists. We orient the graph as follows:
\begin{enumerate}
    \item For each clause, at least one literal evaluates to \texttt{True}. We orient the external edges corresponding to true literals into the clause vertex $v_C$ (out of its variable gadget). The remaining edges are oriented out of $v_C$. Together with the self-loop edge, the total incoming load to $v_C$ is at least $2 +\epsilon > 2$.
    \item If a variable $x$ is \texttt{True}, we orient its cycle counter clockwise. This means internal edges of weight $2$ are oriented from $v^i_{\bar{x}}$ to $v^i_x$, and internal edges of weight $1-\epsilon$ are oriented from $v^{i+1}_{x}$ to $v^{i}_{\bar{x}}$. Furthermore, all external edges representing the negated literal $\bar{x}$ are directed into the cycle (from their respective clauses), while the external edges for $x$ are directed out of the cycle.

    We evaluate the incoming load for any cycle vertex $u$:
    $(i)$ If $u$ is incident to an external edge directed out of the cycle, it is entered by an edge of weight $2$. 
    $(ii)$ If $u$ is incident to an external edge (of weight $1 + \epsilon$) directed  into the cycle, it is entered also by a cycle edge of weight $1-\epsilon$.  The total load into $u$ is $(1+\epsilon) + (1-\epsilon) = 2$.
    $(iii)$ If $u$ is incident to a self loop of weight $1+\epsilon$, then it is entered also by a cycle edge of weight either $1-\epsilon$ or $2$. The total load into $u$ is at least $2$.
    
    If a variable is \texttt{False}, the analysis is symmetric (orienting the cycle clockwise). Thus, the minimum load across all the cycle vertices is exactly $2$.
 \end{enumerate}

For the other direction of the proof, assume there exists an orientation with a minimum load of $2$. 

Consider a variable $x$. Each vertex in a variable gadget of $x$ is incident to a cycle edge of weight $2$, a cycle edge of weight $1-\epsilon$, and an outer edge of weight $1+\epsilon$.  The cycle must be oriented consistently either entirely clockwise or entirely counter-clockwise, as otherwise some vertex would receive a load of at most $1+\epsilon < 2$. 

For any clause vertex $v_C$, the self-loop clearly enters it. To have incoming load at least $2$, at least one of the three incident edges of weight $1+\epsilon$ is oriented into it (out of the variable gadget).

The orientation induces a truth assignment: if a variable's cycle is oriented such that vertices $v_i^x$ are entered by edges of weight $1-\epsilon$, then its positive literal edges must also enter the cycle, and we set it to \texttt{False}; otherwise, we set it to \texttt{True}. Since the cycles are oriented entirely clockwise or counter-clockwise there are no conflicts in the literal assignments, ensuring the resulting truth assignment is perfectly consistent. Since every clause has at least one variable edge into it, every clause is satisfied.
\end{proof}

Next, we show that an oracle that reveals which vertices receive a job of weight strictly greater than $\alpha C_{min}^*$ for $\alpha \le \frac 12$ does not enable an approximation better than $2 -\epsilon$. The weight on the variable gadgets imply that $C_{min}^* \le 2$.

Assume that $C_{min}^*=2$. First, note that every clause vertex $v_C$ is adjacent to four edges, one of weight $1$ and three of weight $1 +\epsilon$. Since $C_{min}^* = 2$, at least one edge of weight $>1$ is entering $v_C$. Since $\alpha C_{min}^* \le 1$ for any $\alpha \le \frac 12$, the oracle will inform all the clause vertices. 

Given that $C_{min}^*=2$, every cycle vertex is entered by either an edge of weight $2$ or an edge of weight $1+\epsilon$. Thus, every cycle vertex has exactly one incoming edge of weight $>1 \ge \alpha C_{min}^*$ regardless of the optimal assignment (and the corresponding truth assignment). The oracle will therefore inform  all variable vertices.
Since this information is independent of the solution, and {\sc $3$-SAT} is NP-hard even if it is known that a truth assignment exists, the oracle does not provide any useful information.

Finally, if $C_{min}^*<2$, then $C_{min}^* \le 1+\epsilon$ implying a gap of $\frac 2{1+\epsilon}$. Since $\epsilon$ can be arbitrarily small, no algorithm can guarantee approximation ratio lower than $2$.
\end{proof}

\begin{theorem}
For any $\frac{1}{2} < \alpha < 1$, it is NP-hard to approximate the {\MMGB} problem within a factor less than $\frac{1}{\alpha}$, even with an oracle that reveals the value of $C_{\min}^*$ and the set of vertices that have an incoming edge of weight more than $\alpha C_{\min}^*$ in some optimal solution.
\end{theorem}

\begin{proof}
Let $0 < \epsilon < 1 - \alpha$ and set $\alpha' = \alpha + \epsilon$, so that $0 < 2 - 2\alpha' < 1 < 2\alpha < 2\alpha' < 2$. We use a similar construction as in the proof of Theorem~\ref{thm:hardSC}. The differences are as follows:
(i) The cycle edges in the variable gadgets alternate in weight between $2$ and $2 - 2\alpha'$.
(ii) All edges connecting a clause vertex and a variable gadget, as well as the self-loops on unmatched variable vertices, have weight $2\alpha'$.
(iii) The self-loop attached to every clause vertex has weight $2 - 2\alpha'$.

Following the proof of Theorem~\ref{thm:hardSC}, the {\MMGB} instance has an orientation with minimum load $C_{min}^* = 2$ if and only if the 3-SAT instance has a satisfying assignment. Also, if $C_{min}^* = 2$, then every clause vertex receives at least one incoming edge of weight $2\alpha' > 2\alpha$, and every variable vertex receives an incoming edge of weight either $2$ or $2\alpha'$ (both strictly greater than $\alpha C_{\min}^* = 2\alpha$). Thus, the oracle informs all vertices in the graph regardless of the optimal orientation, deeming the oracle useless.

Finally, if $C_{\min}^* < 2$, then at least one vertex has incoming load strictly less than $2$, which implies $C_{\min}^* \le \max\{2 - 2\alpha', 2\alpha'\} = 2\alpha'$, yielding a gap of $\frac{2}{2\alpha'} = \frac{1}{\alpha + \epsilon}$. Since $\epsilon > 0$ can be arbitrarily small, no polynomial-time algorithm can guarantee an approximation ratio strictly less than $\frac{1}{\alpha}$.
\end{proof}

\section{Discussion}
\label{sec:discussion}
In this paper we explore the use of trusted advice for classical scheduling problems on unrelated parallel machines. 
Our findings demonstrate that even highly restricted oracle advice can successfully bypass longstanding worst-case approximation barriers. We define several oracles and establish that a single query to our weakest oracle, $\mathsf{Oracle_{LM}}$, which only identifies the subset of machines processing long jobs, improves the approximation of {\MM} from the classical factor $2$ barrier to $\sqrt{3}$. We further refine these techniques to yield improved approximation ratios for other variants of the problem. By duality, we also establish the theoretical limits of trusted advice, showing that these scheduling problems remain NP-hard to approximate and, in some cases, even match their known oracle-free lower bounds.

A natural next step is to further narrow the gaps between our new upper bounds and classical lower bounds. While this work focused on oracles that provide information on the assignment of long jobs, investigating trusted advice that provide different types of information is another challenge. Additionally, applying trusted advice to other problems, such as scheduling with precedence constraints or job due dates is a promising direction. Along these lines, we have already shown that our oracles yield improved approximation ratios for the generalized assignment problem (GAP).

Finally, our algorithms make only a single, highly restricted oracle query. Investigating the trade-off between the number of permitted queries and the resulting approximation guarantees may yield further insight into the power of trusted advice.

\bibliographystyle{plain}  % or use IEEEtran, acm, alpha, etc.
\bibliography{augment}  

\end{document}